\documentclass[11pt]{article}
\usepackage{axodraw}
\usepackage{epsfig}
\usepackage{amsfonts}
\usepackage{amsmath}
\usepackage{bbm}
\usepackage{graphicx}
\allowdisplaybreaks[1]

\input pix.sty

\newcommand{\Li}{\,\mbox{Li}_\rmii{2}}

\newcommand{\ko}{k_0}
\newcommand{\km}{k_-}
\newcommand{\kp}{k_+}
\newcommand{\aL}{a^{ }_\rmii{L}}
\newcommand{\aR}{a^{ }_\rmii{R}}
\renewcommand{\eq}{eq.~}
\renewcommand{\eqs}{eqs.~}
\renewcommand{\se}{sec.~}
\renewcommand{\ses}{secs.~}
\renewcommand{\fig}{fig.~}

\newcommand{\Nc}{N_{\rm c}}

\newcommand{\rmO}{{\mathcal{O}}}

\newcommand{\CA}{\Nc}
\newcommand{\CF}{C_\rmii{F}}

\def\lsi{\raise0.3ex\hbox{$<$\kern-0.75em\raise-1.1ex\hbox{$\sim$}}}
\def\gsi{\raise0.3ex\hbox{$>$\kern-0.75em\raise-1.1ex\hbox{$\sim$}}}
\newcommand{\lsim}{\mathop{\lsi}}
\newcommand{\gsim}{\mathop{\gsi}}

\newcommand{\sign}{\mathop{\mbox{sign}}}
\newcommand{\nF}{n_\rmii{F}}
\newcommand{\nB}{n_\rmii{B}}
 \renewcommand{\nF}[1]{n_\rmii{F{#1}}}
 \renewcommand{\nB}[1]{n_\rmii{B{#1}}}

\newcommand{\rmii}[1]{{\mbox{\tiny\rm{#1}}}}

\newcommand{\re}{\mathop{\mbox{Re}}}
\newcommand{\im}{\mathop{\mbox{Im}}}

\newcommand{\Tint}[1]{{\hbox{$\sum$}\!\!\!\!\!\!\!\int\,}_{\!\!\!\!\raise-0.9ex\hbox{$\scriptstyle{#1}$}}}
\newcommand{\Tinti}[1]{{{\Sigma}\!\!\!\!\raise0.3ex\hbox{$\int$}_\rmii{${#1}$}}}

\newcommand{\bi}{\begin{itemize}}
\newcommand{\ei}{\end{itemize}}

\newcommand{\hide}[1]{ }
\newcommand{\bsl}[1]{\,\slash\!\!\!\!{#1}\,}

\def\TAsc(#1,#2)(#3,#4,#5)%
{\SetWidth{2.0}\CArc(#1,#2)(#3,#4,#5)\SetWidth{1.0}}
\def\Lwidth{3}

\def\TAgl(#1,#2)(#3,#4,#5){\SetWidth{2.0}\PhotonArc(#1,#2)(#3,#4,#5){\Lwidth}%
{6.283 #3 mul 360 div #4 #5 sub #4 #5 sub mul sqrt mul Tdensity mul}%
\SetWidth{1.0}}
\def\TLgl(#1,#2)(#3,#4){\SetWidth{2.0}\Photon(#1,#2)(#3,#4){\Lwidth}
{#1 #3 sub #1 #3 sub mul #2 #4 sub #2 #4 sub mul add sqrt Tdensity mul}%
\SetWidth{1.0}}

\renewcommand{\picc}[1]{\;\parbox[c]{60pt}{\begin{picture}(60,30)(0,0)
\SetWidth{1.0}\SetScale{1.3} #1 \end{picture}}\; }
\def\Lwidth{1.3}

\newcommand{\picv}[1]{\;\parbox[c]{80pt}{\begin{picture}(80,70)(0,-5)
\SetWidth{1.0}\SetScale{1.0} #1 \end{picture}}\; }
\newcommand{\picw}[1]{\;\parbox[c]{160pt}{\begin{picture}(160,70)(-80,-35)
\SetWidth{1.0}\SetScale{1.0} #1 \end{picture}}\; }
\def\Ij{\picv{%
 \Asc(40,30)(30,0,360)%
 \Lqu(-10,30)(10,30)%
 \Lqu(70,30)(90,30)%
 \Photon(40,0)(40,60){1.5}{8}%
 \Text(-3,37)[c]{$\scriptstyle \sigma_0, K$}%
 \Text(83,37)[c]{$\scriptstyle \sigma_0, K$}%
 \Text(60,60)[l]{$\scriptstyle \sigma_2, Q$}%
 \rText(38,30)[l][r]{$\scriptstyle \sigma_5, Q-P$}
 \Text(20,0)[r]{$\scriptstyle \sigma_4, P-K$}%
 \Text(20,60)[r]{$\scriptstyle \sigma_1, P$}%
 \Text(60,0)[l]{$\scriptstyle \sigma_3, Q-K$}%
}}
\def\ronea{\picw{%
 \Lqu(-60,0)(-40,0)%
 \Lqu(40,0)(60,0)%
 \Lsc(-40,0)(-10,30)%
 \Lsc(-40,0)(-10,-30)%
 \Lsc(40,0)(10,30)%
 \Lsc(40,0)(10,-30)%
 \Photon(-25,15)(-10,5){1.5}{4}%
 \Photon(25,-15)(10,-5){1.5}{4}%
 \Text(-50,-7)[c]{$\scriptstyle \sigma_0, \mathcal{K}$}%
 \Text(50,7)[c]{$\scriptstyle \sigma_0, \mathcal{K}$}%
 \Text(-0.5,35)[c]{$\scriptstyle \sigma_2 \; \mathcal{Q}$}%
 \Text(-0.5,0)[c]{$\scriptstyle \sigma_5 \; \mathcal{R}$}
 \Text(-0.5,-35)[c]{$\scriptstyle \sigma_4 \; \mathcal{P}$}%
 \put(-45,5){\mbox{\rotatebox{45}%
    {\hbox{$\scriptstyle \sigma_1, \mathcal{K-P}$}}}}%
 \put(19,-28){\mbox{\rotatebox{45}%
    {\hbox{$\scriptstyle \sigma_3, \mathcal{K-Q}$}}}}%
 \SetWidth{0.5}%
 \Line(0,-35)(0,35)%
}}
\def\roneb{\picw{%
 \Lqu(-60,0)(-40,0)%
 \Lqu(40,0)(60,0)%
 \Lsc(-40,0)(-10,30)%
 \Lsc(-40,0)(-10,-30)%
 \Lsc(40,0)(10,30)%
 \Lsc(40,0)(10,-30)%
 \Photon(-25,-15)(-10,-5){1.5}{4}%
 \Photon(25,15)(10,5){1.5}{4}%
 \Text(-50,7)[c]{$\scriptstyle \sigma_0, \mathcal{K}$}%
 \Text(50,-7)[c]{$\scriptstyle \sigma_0, \mathcal{K}$}%
 \Text(0,35)[c]{$\scriptstyle \sigma_1 \; \mathcal{Q}$}%
 \Text(0,0)[c]{$\scriptstyle \sigma_5 \; \mathcal{R}$}
 \Text(0,-35)[c]{$\scriptstyle \sigma_3 \; \mathcal{P}$}%
 \put(-45,-7){\mbox{\rotatebox{-45}%
    {\hbox{$\scriptstyle \sigma_4, \mathcal{K-Q}$}}}}%
 \put(18,27){\mbox{\rotatebox{-45}%
    {\hbox{$\scriptstyle \sigma_2, \mathcal{K-P}$}}}}%
 \SetWidth{0.5}%
 \Line(0,-35)(0,35)%
}}
\def\rtwoa{\picw{%
 \Lqu(-70,-15)(-40,-15)%
 \Lqu(50,0)(70,-15)%
 \Lsc(-70,15)(-40,15)%
 \Lsc(-40,15)(-40,-15)%
 \Lsc(-40,-15)(-10,-15)%
 \Lsc(10,-15)(30,0)%
 \Lsc(30,0)(50,0)%
 \Lsc(50,0)(70,15)%
 \Photon(-40,15)(-10,15){1.5}{6}%
 \Photon(10,15)(30,0){1.5}{5}%
 \Text(-60,-22)[c]{$\scriptstyle \sigma_0, \mathcal{K}$}%
 \Text(70,-22)[c]{$\scriptstyle \sigma_0, \mathcal{K}$}%
 \Text(-60,22)[c]{$\scriptstyle \sigma_2, \mathcal{Q}$}%
 \Text(70,22)[c]{$\scriptstyle \sigma_2, \mathcal{Q}$}%
 \Text(0,15)[c]{$\scriptstyle \sigma_5 \; \mathcal{R}$}
 \Text(0,-15)[c]{$\scriptstyle \sigma_4 \; \mathcal{P}$}%
 \Text(-18,0)[r]{$\scriptstyle \sigma_1 \; \mathcal{K-P}$}%
 \Text(40,13)[c]{$\scriptstyle \sigma_3, \mathcal{K+Q}$}%
 \SetWidth{0.5}%
 \Line(0,-20)(0,20)%
}}
\def\rtwob{\picw{%
 \Laqu(70,-15)(40,-15)%
 \Laqu(-50,0)(-70,-15)%
 \Lsc(70,15)(40,15)%
 \Lsc(40,15)(40,-15)%
 \Lsc(40,-15)(10,-15)%
 \Lsc(-10,-15)(-30,0)%
 \Lsc(-30,0)(-50,0)%
 \Lsc(-50,0)(-70,15)%
 \Photon(40,15)(10,15){1.5}{6}%
 \Photon(-10,15)(-30,0){1.5}{5}%
 \Text(60,-22)[c]{$\scriptstyle \sigma_0, \mathcal{K}$}%
 \Text(-70,-22)[c]{$\scriptstyle \sigma_0, \mathcal{K}$}%
 \Text(60,22)[c]{$\scriptstyle \sigma_1, \mathcal{Q}$}%
 \Text(-70,22)[c]{$\scriptstyle \sigma_1, \mathcal{Q}$}%
 \Text(0,15)[c]{$\scriptstyle \sigma_5 \; \mathcal{R}$}
 \Text(0,-15)[c]{$\scriptstyle \sigma_3 \; \mathcal{P}$}%
 \Text(30,0)[l]{$\scriptstyle \sigma_2 \; \mathcal{K-P}$}%
 \Text(-40,13)[c]{$\scriptstyle \sigma_4, \mathcal{K+Q}$}%
 \SetWidth{0.5}%
 \Line(0,-20)(0,20)%
}}
\def\rthreea{\picw{%
 \Lqu(-70,-15)(-40,-15)%
 \Lqu(40,15)(70,15)%
 \Lsc(-40,15)(-10,15)%
 \Lsc(-40,15)(-40,-15)%
 \Lsc(-40,-15)(-10,-15)%
 \Lsc(40,15)(10,15)%
 \Lsc(40,15)(40,-15)%
 \Lsc(40,-15)(10,-15)%
 \Photon(-70,15)(-40,15){1.5}{6}%
 \Photon(40,-15)(70,-15){1.5}{6}%
 \Text(-70,-22)[c]{$\scriptstyle \sigma_0, \mathcal{K}$}%
 \Text(70,22)[c]{$\scriptstyle \sigma_0, \mathcal{K}$}%
 \Text(-70,22)[c]{$\scriptstyle \sigma_5, \mathcal{R}$}%
 \Text(70,-22)[c]{$\scriptstyle \sigma_5, \mathcal{R}$}%
 \Text(0,15)[c]{$\scriptstyle \sigma_2 \; \mathcal{Q}$}
 \Text(0,-15)[c]{$\scriptstyle \sigma_4 \; \mathcal{P}$}%
 \Text(-18,0)[r]{$\scriptstyle \sigma_1 \; \mathcal{K-P}$}%
 \Text(30,0)[l]{$\scriptstyle \sigma_3 \; \mathcal{K-Q}$}%
 \SetWidth{0.5}%
 \Line(0,-20)(0,20)%
}}
\def\rthreeb{\picw{%
 \Laqu(70,-15)(40,-15)%
 \Laqu(-40,15)(-70,15)%
 \Lsc(40,15)(10,15)%
 \Lsc(40,15)(40,-15)%
 \Lsc(40,-15)(10,-15)%
 \Lsc(-40,15)(-10,15)%
 \Lsc(-40,15)(-40,-15)%
 \Lsc(-40,-15)(-10,-15)%
 \Photon(70,15)(40,15){1.5}{6}%
 \Photon(-40,-15)(-70,-15){1.5}{6}%
 \Text(70,-22)[c]{$\scriptstyle \sigma_0, \mathcal{K}$}%
 \Text(-70,22)[c]{$\scriptstyle \sigma_0, \mathcal{K}$}%
 \Text(70,22)[c]{$\scriptstyle \sigma_5, \mathcal{R}$}%
 \Text(-70,-22)[c]{$\scriptstyle \sigma_5, \mathcal{R}$}%
 \Text(0,15)[c]{$\scriptstyle \sigma_1 \; \mathcal{Q}$}
 \Text(0,-15)[c]{$\scriptstyle \sigma_3 \; \mathcal{P}$}%
 \Text(30,0)[l]{$\scriptstyle \sigma_2 \; \mathcal{K-P}$}%
 \Text(-18,0)[r]{$\scriptstyle \sigma_4 \; \mathcal{K-Q}$}%
 \SetWidth{0.5}%
 \Line(0,-20)(0,20)%
}}
\def\rfoura{\picw{%
 \Laqu(70,15)(40,15)%
 \Laqu(-50,0)(-70,15)%
 \Lsc(70,-15)(40,-15)%
 \Lsc(40,-15)(40,15)%
 \Lsc(40,15)(10,15)%
 \Lsc(-10,15)(-30,0)%
 \Lsc(-30,0)(-50,0)%
 \Lsc(-50,0)(-70,-15)%
 \Photon(40,-15)(10,-15){1.5}{6}%
 \Photon(-10,-15)(-30,0){1.5}{5}%
 \Text(60,22)[c]{$\scriptstyle \sigma_0, \mathcal{K}$}%
 \Text(-70,22)[c]{$\scriptstyle \sigma_0, \mathcal{K}$}%
 \Text(60,-22)[c]{$\scriptstyle \sigma_4, \mathcal{P}$}%
 \Text(-70,-22)[c]{$\scriptstyle \sigma_4, \mathcal{P}$}%
 \Text(0,-15)[c]{$\scriptstyle \sigma_5 \; \mathcal{R}$}
 \Text(0,15)[c]{$\scriptstyle \sigma_2 \; \mathcal{Q}$}%
 \Text(30,0)[l]{$\scriptstyle \sigma_3 \; \mathcal{K-Q}$}%
 \Text(-40,-13)[c]{$\scriptstyle \sigma_1, \mathcal{K+P}$}%
 \SetWidth{0.5}%
 \Line(0,-20)(0,20)%
}}
\def\rfourb{\picw{%
 \Lqu(-70,15)(-40,15)%
 \Lqu(50,0)(70,15)%
 \Lsc(-70,-15)(-40,-15)%
 \Lsc(-40,-15)(-40,15)%
 \Lsc(-40,15)(-10,15)%
 \Lsc(10,15)(30,0)%
 \Lsc(30,0)(50,0)%
 \Lsc(50,0)(70,-15)%
 \Photon(-40,-15)(-10,-15){1.5}{6}%
 \Photon(10,-15)(30,0){1.5}{5}%
 \Text(-60,22)[c]{$\scriptstyle \sigma_0, \mathcal{K}$}%
 \Text(70,22)[c]{$\scriptstyle \sigma_0, \mathcal{K}$}%
 \Text(-60,-22)[c]{$\scriptstyle \sigma_3, \mathcal{P}$}%
 \Text(70,-22)[c]{$\scriptstyle \sigma_3, \mathcal{P}$}%
 \Text(0,-15)[c]{$\scriptstyle \sigma_5 \; \mathcal{R}$}
 \Text(0,15)[c]{$\scriptstyle \sigma_1 \; \mathcal{Q}$}%
 \Text(-18,0)[r]{$\scriptstyle \sigma_4 \; \mathcal{K-Q}$}%
 \Text(40,-13)[c]{$\scriptstyle \sigma_2, \mathcal{K+P}$}%
 \SetWidth{0.5}%
 \Line(0,-20)(0,20)%
}}
\def\va{\picw{%
 \Lqu(-60,0)(-40,0)%
 \Lqu(50,0)(70,0)%
 \Lsc(-40,0)(-20,15)%
 \Lsc(-40,0)(-20,-15)%
 \Lsc(0,18)(25,18)%
 \Lsc(0,-18)(25,-18)%
 \Lsc(25,18)(50,0)%
 \Lsc(25,-18)(50,0)%
 \Photon(25,-18)(25,18){-1.5}{6.5}%
 \Text(-60,7)[c]{$\scriptstyle \sigma_0, \mathcal{K}$}%
 \Text(70,7)[c]{$\scriptstyle \sigma_0, \mathcal{K}$}%
 \Text(-4,-24)[c]{$\scriptstyle \sigma_4 , \mathcal{K-P}$}
 \Text(-10,20)[c]{$\scriptstyle \sigma_1 \; \mathcal{P}$}%
 \rText(23,0)[r][r]{$\scriptstyle \sigma_5, \mathcal{P-Q}$}%
 \Text(35,19)[l]{$\scriptstyle \sigma_2, \mathcal{Q}$}%
 \Text(35,-19)[l]{$\scriptstyle \sigma_3, \mathcal{K-Q}$}%
 \SetWidth{0.5}%
 \Line(-10,-20)(-10,20)%
}}
\def\vb{\picw{%
 \Laqu(60,0)(40,0)%
 \Laqu(-50,0)(-70,0)%
 \Lsc(40,0)(20,15)%
 \Lsc(40,0)(20,-15)%
 \Lsc(0,18)(-25,18)%
 \Lsc(0,-18)(-25,-18)%
 \Lsc(-25,18)(-50,0)%
 \Lsc(-25,-18)(-50,0)%
 \Photon(-25,-18)(-25,18){1.5}{6.5}%
 \Text(60,7)[c]{$\scriptstyle \sigma_0, \mathcal{K}$}%
 \Text(-70,7)[c]{$\scriptstyle \sigma_0, \mathcal{K}$}%
 \Text(16,-24)[c]{$\scriptstyle \sigma_3 , \mathcal{K-P}$}
 \Text(9.5,21)[c]{$\scriptstyle \sigma_2 \; \mathcal{P}$}%
 \rText(-24,0)[l][l]{$\scriptstyle \sigma_5, \mathcal{P-Q}$}%
 \Text(-35,19)[r]{$\scriptstyle \sigma_1, \mathcal{Q}$}%
 \Text(-35,-19)[r]{$\scriptstyle \sigma_4, \mathcal{K-Q}$}%
 \SetWidth{0.5}%
 \Line(10,-20)(10,20)%
}}
\makeatletter \@addtoreset{equation}{section} \makeatother
\renewcommand{\theequation}{\arabic{section}.\arabic{equation}}
\makeatletter
\renewcommand\section{\@startsection {section}{1}{\z@}%
                                   {-5.5ex \@plus -1ex \@minus -.2ex}
                                   {2.3ex \@plus.2ex}%
                                   {\normalfont\large\bfseries}}
\renewcommand\subsection{\@startsection{subsection}{2}{\z@}%
                                     {-3.25ex\@plus -1ex \@minus -.2ex}%
                                     {1.5ex \@plus .2ex}%
                                     {\normalfont\normalsize\bfseries}}
\renewcommand\thesection {\@arabic\c@section}
\renewcommand\thesubsection   {\thesection.\@arabic\c@subsection}
\renewcommand{\@seccntformat}[1]{%
\csname the#1\endcsname.\hspace{1.0em}}
\makeatother

\begin{document}

\flushbottom

\begin{titlepage}

\begin{flushright}
arXiv:1304.0202\\ 
\end{flushright}

\begin{centering}
\vfill

{\Large{\bf
 Thermal 2-loop master spectral function at finite momentum
 }}

\vspace{0.8cm}

M.~Laine 

\vspace{0.8cm}

{\em
ITP, AEC, University of Bern, 
Sidlerstrasse 5, CH-3012 Bern, Switzerland\\}

\vspace*{0.8cm}

\mbox{\bf Abstract}

\end{centering}
  
\vspace*{0.3cm}
 
\noindent
%
When considering NLO corrections to thermal particle production in the
``relativistic'' regime, in which the invariant mass squared of the produced
particle is $\mathcal{K}^2 \sim (\pi T)^2$, then the production rate can 
be expressed as a sum of a few universal ``master'' spectral
functions. Taking the most complicated 2-loop master as an example, 
a general strategy for obtaining a convergent 2-dimensional integral 
representation is suggested. The analysis applies both to bosonic and 
fermionic statistics, and shows that for this master
the non-relativistic approximation is only accurate for $\mathcal{K}^2 \gsim 
(8 \pi T)^2$, whereas the zero-momentum approximation works surprisingly well.
Once the simpler masters have been similarly resolved, NLO 
results for quantities such as the right-handed neutrino production rate 
from a Standard Model plasma or the dilepton production rate from a QCD 
plasma can be assembled for $\mathcal{K}^2 \sim (\pi T)^2$.
%

\vfill

 
\vspace*{1cm}
  
\noindent
May 2013

\vfill
  
\end{titlepage}


%
\section{Introduction}

Given the remarkable confirmation of microscopic Standard Model 
and QCD physics through the Large Hadron Collider 
program, it appears well-motivated to work out  
how the same interactions behaved in the macroscopic 
thermal environment of the Early Universe. Whereas there are good 
tools available for doing this for thermodynamic quantities such
as the overall equation of state, much less is known about real-time
rates, such as particle production rates or the equilibration rates
related to the most weakly interacting degrees of freedom. Indeed, 
in most cases only phenomenological estimates or 
leading-order (LO) weak-coupling expressions
are available. Given the known infrared problems of thermal field
theory, which imply that 
next-to-leading order (NLO) corrections may be surprisingly 
large, further work is needed in order to get an impression 
on the numerical accuracy of the results currently available.  

In relativistic thermal field theory, the structure of the weak-coupling
expansion depends sensitively on the physical scales of 
the problem. Focussing on the simplest situation, in which there is only 
one zero-temperature mass scale, denoted by the invariant mass squared
$\mathcal{K}^2 > 0$, there are different ``regimes'' depending 
on the ratio $\mathcal{K}^2 / (\pi T)^2$.\footnote{%
 To be precise, because of the absence of Lorentz invariance
 within a heat bath  
 the magnitudes of $k^{}_\pm \equiv (\ko\pm k)/2$ need to 
 be separately compared with $\pi T$, and the number of 
 regimes may proliferate.
 } 
In the so-called {\em non-relativistic} regime, 
$\mathcal{K}^2 \gg (\pi T)^2$, results can be represented 
in the form of an Operator Product Expansion~\cite{simon}, 
and thermal corrections are in general small (power-suppressed). 
In contrast, in the so-called {\em ultrarelativistic} regime, 
$\mathcal{K}^2 \ll (\pi T)^2$, the naive loop expansion 
breaks down, and extensive resummations (incorporating for instance
the physics of the Landau-Pomeranchuk-Migdal effect) 
are needed for determining even the LO result. The current 
technology for this has been developed in the context of
the photon production rate from 
a QCD plasma~\cite{amypre,amy,amypost}\footnote{%
 Similar resummations are needed for dilepton production in 
 the soft regime $\mathcal{K}^2 \ll (\pi T)^2$~\cite{bp,ag,mr}.
 }
and subsequently reformulated 
and applied to the right-handed neutrino production rate
from a Standard Model plasma~\cite{bb0,bb1,bb2}.
(Very recently the photon production rate 
has been determined up to NLO~\cite{photonNLO}, which here 
means $\rmO(\alpha^{1/2})$ rather than the usual $\rmO(\alpha)$, 
where $\alpha \equiv g^2/(4\pi)$.)

The focus of the present paper is the so-called 
{\em relativistic} regime, $\mathcal{K}^2 \sim (\pi T)^2$.
Such results may permit for an interpolation
between the non-relativistic and ultrarelativistic cases, 
thereby perhaps yielding phenomenologically broadly
applicable expressions (the cosmological evolution
is illustrated in \fig\ref{fig:evolution}). 
At NLO, which in the relativistic 
regime means $\rmO(\alpha)$, no infrared divergences are
expected to be encountered in the full result, 
and there is no need for resummations 
either. However, because of the loss of Lorentz symmetry within 
a thermal bath, the analysis is technically cumbersome.  This is the 
case particularly if a non-zero momentum $k \equiv |\vec{k}| \neq 0$
is considered, in which case even LO expressions are 
non-trivial~\cite{aarts}. Nevertheless, as will be demonstrated, 
NLO results can be worked out with some effort. 

\begin{figure}[t]


\centerline{%
 \epsfysize=7.5cm\epsfbox{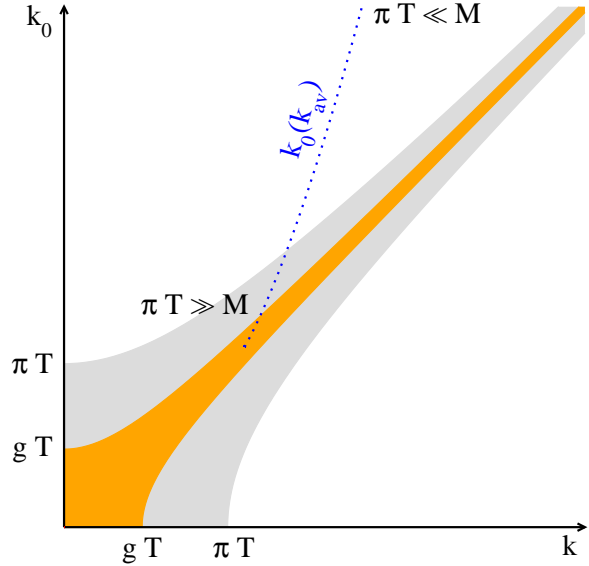}%
}

\caption[a]{\small
An example of how during the cosmological evolution, different 
kinematic regimes are crossed in the thermal production of a particle 
species with a fixed invariant mass $\mathcal{K}^2 \equiv M^2$. A generic
weak gauge coupling is denoted by $g$ [$\alpha\equiv g^2/(4\pi)$], and
$k^{ }_\rmi{av}$ is defined by \eq\nr{kav}. If a 
fixed invariant energy 
$(E^{ }_{\mu^-} + E^{ }_{\mu^+})^2 - 
(\vec{k}^{ }_{\mu^-} + \vec{k}^{ }_{\mu^+})^2 \simeq 1$~GeV$^2$
of a dilepton pair is considered, a similar
evolution takes place during a heavy ion 
collision, however the temperature range is rather narrow. 
}

\la{fig:evolution}
\end{figure}

Examples of 
concrete applications that we have in mind are the right-handed
neutrino production rate in the Early Universe, 
motivated by leptogenesis or dark matter 
computations~\cite{yanagida,numsm}, and the dilepton 
production rate in hot QCD, motivated by 
heavy ion collision experiments. 
For the former, the NLO level has already
been reached in the non-relativistic regime~\cite{salvio,nonrel}.
For the latter, the NLO level was reached long ago at 
{\em vanishing momentum} 
in the relativistic regime~\cite{spectral1,spectral2,spectral3}.
More recently, the case of a vanishing momentum in the relativistic
regime has been studied more generally~\cite{bulk_wdep}, 
expressing results in terms of a finite number of universal
``master'' structures; in particle language 
the cases considered were the (off-shell) thermal 
axion and dilaton~\cite{bulk_wdep}
or graviton~\cite{shear_wdep} production rates.
(The physical relevance of these computations
is related to lattice determinations of the corresponding ``transport
coefficients'', which are the rate of anomalous chirality violation 
and the bulk and shear viscosities. Indeed, because of the necessity 
of analytic continuation from Euclidean signature, 
ultraviolet features need to subtracted from non-perturbative data 
before an extrapolation to $\ko \to 0$ becomes 
possible even in principle~\cite{cuniberti,analytic,cond}.)
 
The purpose of the present paper is to evaluate 
the {\em spectral function} 
corresponding to the most complicated 2-loop master
topology (\fig\ref{fig:Ij}) at non-zero momentum
in the relativistic regime. 
This topology has  
merited extensive investigations under other circumstances. 
Indeed, at zero temperature in 3 space-time dimensions
even the massive case can be solved~\cite{akr}, whereas in the physical
4-dimensional case the result has an extremely rich structure~\cite{db}
which, despite a vast body of work, 
still remains under further investigation today 
(see, e.g., ref.~\cite{vs} and references therein). 
In the massless limit, however, the result 
vanishes at zero temperature. Nevertheless the spectral function
(cut) is non-zero and possesses 
a rich structure at finite temperature. Previous 
thermal analyses exist at zero momentum~\cite{bulk_wdep}
as well as at zero energy but non-zero momentum~\cite{bulk_rdep}, 
the latter case corresponding to the space-like domain relevant for 
the physics of plasma screening.

Recently, a separate line of study of thermal particle 
production in the relativistic regime
has been initiated whose goals
appear to partly overlap with those of 
the present paper~\cite{gh}. The authors give a 4-dimensional 
integral representation for the gauge boson contribution to the right-handed
neutrino production rate, but no numerical evaluation is shown. Given the 
technical complexity of the problem, it appears welcome that two 
independent and methodologically different computations are being pursued, 
permitting in the end to crosscheck the validity of both results.  

The plan of this paper is the following. We start by recalling how NLO 
results for two physical observables can be expressed in terms
of a finite number of simple master sum-integrals (\se\ref{se:reduce}). 
After carrying out Matsubara sums, the cut of the most complicated master can 
furthermore be decomposed into processes representing real and virtual
corrections (\se\ref{se:split}). The real corrections are analyzed
in \se\ref{se:real}, showing that through a suitable choice of 
variables they can be given a 2-dimensional integral representation
(soft and collinear divergences are regulated by an auxiliary mass
parameter at this stage); the same task is accomplished for the 
virtual corrections in \se\ref{se:virtual}. Both real and virtual 
corrections are divergent if the auxiliary mass parameter is sent 
to zero; in \se\ref{se:sum} it is shown that the sum remains finite. 
Finally a form suitable for practical evaluation is given 
in \se\ref{se:final}, whereas numerical
comparisons with known limiting values comprise
\se\ref{se:num}. A brief summary 
and outlook are offered in \se\ref{se:concl}.

%
\section{Physical observables}
\la{se:reduce}

To underline the significance of 
the master spectral function considered, we start
by recalling two separate physics contexts in which it plays a role. 

To leading order in $\alpha^{ }_{em} \equiv e^2/(4\pi)$, 
the production rate of $\mu^-\mu^+$ pairs from a hot QCD medium
can be expressed as~\cite{dilepton1,dilepton2,dilepton3}
\be
 \frac{{\rm d} N_{\mu^-\mu^+}}
   {{\rm d}^4 \mathcal{X} {\rm d}^4 \mathcal{K}} =  
 \sum_{qq'}
 \frac{ -2 e^4 Q_{q} Q_{q'} \, \theta(\mathcal{K}^2 - 4 m_\mu^2) } 
  {3 (2\pi)^5 \mathcal{K}^2} 
 \biggl( 1 + \frac{2 m_\mu^2}{\mathcal{K}^2}
 \biggr)
 \biggl(
 1 - \frac{4 m_\mu^2}{\mathcal{K}^2} 
 \biggr)^\fr12 \nB{} (\ko) \rho_{qq'}(\mathcal{K})
 \;, 
 \la{rate_dilepton} 
\ee
where $Q_{q}$ is the electric charge of quark of flavour $q$  
in units of $e$, $\nB{}$ is the Bose distribution, and 
\be
  \rho_{qq'}(\mathcal{K}) \equiv 
  \int_\mathcal{X} 
   e^{i \mathcal{K}\cdot \mathcal{X}}
  \left\langle
    \fr12 \bigl[ 
    \hat{\mathcal{J}}_{q}^\mu (\mathcal{X}), 
    \hat{\mathcal{J}}^{ }_{q'\mu}(0)
    \bigr]
  \right\rangle^{ }_{T}
\ee 
is the spectral function corresponding to the vector 
current. (The metric convention 
$\mathcal{K}\cdot\mathcal{X} = \ko x_0 - \vec{k}\cdot\vec{x}$ is assumed.)
The spectral function can in turn be expressed as the cut of 
the corresponding Euclidean correlator, 
\be
 \rho_{qq'}(\mathcal{K}) = 
 \im \Bigl[ \Pi_{\rmii{E}qq'}(K) \Bigr]^{ }_{k_n \to -i [\ko + i 0^+]}
 \;. \la{cut_dilepton}
\ee
Here $K = (k_n,\vec{k})$ denotes a Matsubara four-momentum, with 
$K^2 = k_n^2 + k^2$, $k \equiv |\vec{k}|$.

The Euclidean correlator can be computed with standard
path integral techniques. 
After carrying out the Dirac algebra and making use of substitutions of
integration variables, it can be expressed as a sum of a finite number of
``master'' structures. Up to 2-loop level, the expression reads
\ba
 \Pi_{\rmii{E}qq'}(K) \!\! & = & \!\! 
 2 (D-2) \CA \delta^{ }_{qq'}
 \Tint{\{P\}} 
 \biggl[ \frac{K^2}{P^2(P-K)^2} - \frac{2}{P^2} \biggr]
 \nn 
 & + & \!\! 4 (D-2) g^2 \CA \CF \delta^{ }_{qq'} 
 \biggl\{ 
   \Tint{ \{P\}Q }
   \biggl[
     \frac{D-2}{Q^2 P^4} - \frac{2}{Q^2P^2(P-K)^2}
   - \frac{(D-2)K^2}{Q^2P^4(P-K)^2} 
   \biggr]
  \nn & & + \, 
   \Tint{ \{PQ\} }
  \biggl[
    - \frac{D-2}{Q^2 P^4}
    + \frac{D-4}{Q^2P^2(Q-P)^2}
    + \frac{2}{Q^2P^2(P-K)^2}
    + \frac{(D-2)K^2}{Q^2P^4(P-K)^2}
  \nn & & \quad - \,
  \frac{D-4}{Q^2(Q-P)^2 (P-K)^2} + 
  \frac{\fr12 (D-7) K^2}{P^2(P-K)^2Q^2(Q-K)^2}
  \nn[2mm] & & \quad + \, 
  \frac{2(D-2)K\cdot Q- (D-6) K^2}{Q^2 P^2 (Q-P)^2 (P-K)^2} - 
  \frac{K^4}{Q^2 P^2 (Q-P)^2 (Q-K)^2 (P-K)^2 }
  \biggr] 
 \biggr\} 
 \;. \nn \la{dilepton}
\ea
Here $D$ is the dimensionality of space-time, 
and $\CF \equiv (\Nc^2-1)/(2\Nc)$. According to standard conventions, 
$\Tinti{\{ P \}}$ denotes a sum-integral with fermionic Matsubara 
momenta, and $\Tinti{ P }$ one with bosonic Matsubara momenta. 
(The generalization of \eq\nr{dilepton} to a finite quark mass
can be found in ref.~\cite{GVtau}.)

%
\begin{figure}[t]
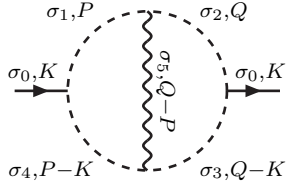


\begin{eqnarray*}
&& 
 \hspace*{-1cm}
 \Ij
\end{eqnarray*}

\caption[a]{\small 
The master Feynman diagram considered, defined by \eq\nr{def_Ij}. 
The integration measure corresponds
to a Matsubara sum-integral, and the spectral function is the 
discontinuity (imaginary part) of the result 
once the external Matsubara frequency
is analytically continued as $k_n \to -i [\ko + i 0^+]$. 
Dashed lines represent massless propagators; the wiggly line 
is regulated through a small mass parameter, $\lambda$, 
for intermediate
stages of the computation. The internal and external lines could be 
either fermions or bosons [denoted by $\sigma^{ }_\mu = -1$ and $+1$, 
respectively, with $\mu\in \{ 0,...,5\}$], 
however fermion number conservation is 
assumed at each vertex. 
} 
\la{fig:Ij}
\end{figure}
%

The different terms of \eq\nr{dilepton} can be referred to 
as master sum-integrals; their cuts, defined in accordance with 
\eq\nr{cut_dilepton}, are the corresponding master spectral functions. 
In particular, the most complicated master sum-integral 
(the only one with 5 different propagators) is defined from now on as 
\ba
 \mathcal{I}^{ }_\rmi{j}(K) & \!\!\equiv\!\! & 
 \lim^{ }_{\lambda \to 0}
 \Tint{PQ} \frac{K^4}{Q^2P^2[(Q-P)^2+\lambda^2](Q-K)^2(P-K)^2} 
 \;,  \la{def_Ij} 
\ea
and is illustrated graphically in \fig\ref{fig:Ij}. Note that a mass
regulator $\lambda$ has been introduced for intermediate stages of 
the computation. The subscript $(...)^{ }_\rmii{j}$ corresponds to 
the labelling of the various masters in ref.~\cite{nonrel}.
The statistics
of $P,Q$ are meant to be interpreted openly, and are labelled by the 
indices $\sigma_0 ... \sigma_5$ as shown in \fig\ref{fig:Ij}. 

The indices $\sigma_0 ... \sigma_5$ 
take the value $+1$ for bosons and $-1$ for fermions. 
Assuming fermion number conservation at each vertex, only three 
of the indices are independent; 
we can choose $\sigma_1, \sigma_4, \sigma_5$ to play 
this role. Then 
\be
 \sigma_0 = \sigma_1 \sigma_4  \;; \quad
 \sigma_2 = \sigma_1 \sigma_5 \;; \quad
 \sigma_3 = \sigma_4 \sigma_5 \;.  \la{i_ids}
\ee
The case appearing in \eq\nr{dilepton} corresponds to 
$(\sigma_1 \sigma_4 \sigma_5) = $($-$\,$-$\,$+$), 
and a numerical evaluation
for the corresponding spectral function, 
defined in \eq\nr{cut}, is shown 
in \fig\ref{fig:ope} below
[along the curve in the ($k,\ko$)-plane shown 
in \fig\ref{fig:evolution}]. 

A completely different physics application, leading to the same 
master spectral functions but with (partly) different statistics, 
is that of right-handed neutrino production in the Early Universe. 
Again, a gauge-invariant Euclidean correlator can be defined, 
\be
 \Pi_\rmii{E}^{ }(K) \equiv 
 \tr\Bigl\{ 
   i \bsl{K}  
   \Bigl[
     \int_0^{1/T} \! {\rm d}\tau \int_\vec{x} e^{i K\cdot X}
   \Bigl\langle
     (\tilde{\phi}^\dagger \aL \ell)(X) \; 
     (\bar{\ell}\,  \aR \tilde\phi)(0)
   \Bigr\rangle^{ }_T 
   \Bigr]
 \Bigr\}
 \;, \la{PiE_def}
\ee
where $X \equiv (\tau,\vec{x})$; 
$\ell$ is a lepton doublet; $\tilde\phi$ 
is a Higgs doublet; and $\aL$, $\aR$ are chiral projectors. 
The production rate now reads 
\be
 \frac{{\rm d}N^{ }_{\nu^{ }_\rmii{R}}
  (\mathcal{K})}{{\rm d}^4\mathcal{X} {\rm d}^3\vec{k}}
 = \frac{2 |h^{ }_\nu|^2}{(2\pi)^3 \ko } 
 \nF{}(\ko) \rho(\mathcal{K})
 \;, \la{rate}
\ee
where $\nF{}$ is the Fermi distribution; $h_\nu^{ }$ a bare 
neutrino Yukawa coupling; and $\ko = \sqrt{k^2 + M^2}$, with $M$
denoting the right-handed neutrino mass. The spectral function $\rho$
is obtained from the cut of 
the Euclidean correlator $\Pi^{ }_\rmii{E}$ just 
like in \eq\nr{cut_dilepton}.

The 2-loop expression for $\Pi^{ }_\rmii{E}$ can be written 
in a form analogous to \eq\nr{dilepton}~\cite{nonrel}:
\ba
 && \hspace*{-1cm} \Pi^{ }_\rmii{E}(K)  =  
  - \Tint{ P }
 \biggl[ 
   \frac{2}{P^2} + \frac{2K^2}{P^2(P-K)^2}
 \biggr]
  + \Tint{ \{ P \} }
  \frac{2}{P^2}
 \nn & + & 
 12 \lambda_h \Tint{ PQ } 
 \biggl[
   \frac{1}{Q^2 P^4} - \frac{1}{Q^2 P^2 (P-K)^2 } 
   + \frac{K^2}{Q^2 P^4 (P-K)^2} 
 \biggr] 
 \nn & + & 
 2 \Nc |h_t|^2 
 \Tint{ P \{ Q\} }
 \biggl[
   -\frac{2}{Q^2 P^4} + \frac{1}{Q^2 P^2 (Q-P)^2} + \frac{2}{Q^2 P^2 (P-K)^2}
  - \frac{2 K^2}{Q^2 P^4 (P-K)^2}
 \nn &  & \quad - \, 
   \frac{1}{Q^2(Q-P)^2(P-K)^2} + 
  \frac{K^2}{Q^2 P^2 (Q-P)^2(P-K)^2}
  \biggr]
 \nn[2mm] & + &
 \frac{g_1^2 + 3 g_2^2}{2}
 \biggl\{ 
   \Tint{ PQ } \biggl[ 
     \frac{D-1}{Q^2 P^4} - \frac{2}{Q^2 P^2 (Q-P)^2}
   - \frac{1}{Q^2 P^2(P-K)^2} + \frac{(D-1) K^2}{Q^2 P^4 (P-K)^2}
 \nn &  & \quad + \, 
  \frac{2 K^2}{P^2(P-K)^2 Q^2 (Q-K)^2} - \frac{4 K^2}{Q^2 P^2 (Q-P)^2(P-K)^2}
 \nn &  & \quad + \, 
   \frac{2 K^4}{Q^2 P^2 (Q-P)^2 (Q-K)^2(P-K)^2}
   \biggr]
 \nn &  & + \, 
 \Tint{ \{P\}Q }
 \biggl[
   - \frac{D-2}{Q^2 P^4} 
   + \frac{(D-2)K^2}{Q^2 P^4(P-K)^2}
   + \frac{2(D-2)K\cdot Q - 4 K^2}{Q^2 P^2(Q-P)^2 (P-K)^2} 
 \biggr]
 \nn[2mm] &  & + \, 
 \Tint{ P\{Q\} }
 \biggl[
   \frac{2}{Q^2 P^2(Q-P)^2} - \frac{D-2}{Q^2 P^2(P-K)^2} 
 \biggr]
 \nn &  & + \, 
  \Tint{ \{PQ\} }
 \biggl[
   \frac{D-2}{Q^2 P^4} - \frac{(D-2)K^2}{Q^2 P^4(P-K)^2} 
 \biggr]
 \biggr\}
 \;. \la{neutrino}
\ea
Here $\lambda_h$ is the Higgs self-coupling, $h_t$ is the top Yukawa 
coupling, and $g^{ }_1$, $g^{ }_2$ are the hypercharge and weak 
gauge coupling, respectively. 

The master sum-integrals appearing in \eq\nr{neutrino} 
are identical to those in \eq\nr{dilepton}, apart from 
their statistics. Noting that $K$ is now fermionic, 
the most complicated case, \eq\nr{def_Ij}, 
appears with the indices
$(\sigma_1 \sigma_4 \sigma_5)$ = ($+$\,$-$\,$+$).
A numerical evaluation is presented in \fig\ref{fig:ope}
along the curve in the ($k,\ko$)-plane shown 
in \fig\ref{fig:evolution}. 

The two examples discussed should serve as illustrations of 
concrete applications of the spectral function related to 
\eq\nr{def_Ij}, but do not exhaust the cases considered
in the literature. In particular, physical observables leading 
to the case 
$(\sigma_1 \sigma_4 \sigma_5)$ = ($+$\,$+$\,$+$) have 
been analyzed in some detail for $k=0$ in the relativistic regime 
in refs.~\cite{bulk_wdep,shear_wdep}, and for $k\neq 0$ in 
the non-relativistic regime in refs.~\cite{bulk_ope,shear_ope,kit}.

As a final remark we mention that above 
the physical observable
was a Lorentz scalar, and subsequently the master structures 
are scalars as well. In the right-handed neutrino case it may 
be of interest to ultimately compute the whole self-energy 
matrix. In this case tensor sum-integrals appear apart from 
scalar ones; the corresponding results have 
been worked out in the non-relativistic regime in ref.~\cite{selfE}.
A general discussion concerning the thermal tensor basis 
can be found in ref.~\cite{hw}.

%
\section{Splitup into real and virtual corrections}
\la{se:split}

Given \eq\nr{def_Ij}, 
the corresponding spectral function can be defined as 
\be
 \rho^{ }_{\mathcal{I}^{ }_\rmii{j}} \equiv
 \im\{ \mathcal{I}^{ }_\rmi{j} \}^{ }_{k_n \to -i [\ko + i 0^+]}
 \;. \la{cut}
\ee
It is well established at 1-loop level that the cut yields 
a structure reminiscent of a Boltzmann equation, with scattering
amplitudes squared multiplied by appropriate phase space
distributions~\cite{weldon}. A similar result applies
at 2-loop level, but is somewhat complicated by the fact that
now virtual particles also appear. In any case, 
making use of standard techniques, explained
in some detail for instance in appendix~A of ref.~\cite{nlo},
and shifting four-momenta, the following result can be obtained: 
\ba
 \frac{2 \rho^{ }_{\mathcal{I}^{ }_\rmii{j}}(\mathcal{K})}
 {\mathcal{K}^4} \! & = & 
 \int_{\vec{p,q,r}}
 \frac{(2\pi)^4 \delta^{(4)}(-\mathcal{K}+\mathcal{P+Q+R})} 
 {8 \epsilon^{ }_p \epsilon^{ }_q E^{ }_r}
 \nn & & \quad \times \, \mathbbm{P} \, \biggl\{ 
 \frac{
  [1 + n^{ }_{\sigma_4}(\epsilon^{ }_p)]
  [1 + n^{ }_{\sigma_2}(\epsilon^{ }_q)]
  [1 + n^{ }_{\sigma_5}(E^{ }_r)] - 
  n^{ }_{\sigma_4}(\epsilon^{ }_p) n^{ }_{\sigma_2}(\epsilon^{ }_q) n^{ }_{\sigma_5}(E^{ }_r)
 }{(\mathcal{K-P})^2 (\mathcal{K-Q})^2 }
 \biggr\}
 \hspace*{0.29cm} \mbox{(r1)} \nn 
 & + & 
 \int_{\vec{p,q,r}}
 \frac{(2\pi)^4 \delta^{(4)}(-\mathcal{K}-\mathcal{P+Q+R})} 
 {8 \epsilon^{ }_p \epsilon^{ }_q E^{ }_r}
 \nn & & \quad \times \, \mathbbm{P} \, \biggl\{ 
 \frac{
  n^{ }_{\sigma_4}(\epsilon^{ }_p) 
  [1 + n^{ }_{\sigma_2}(\epsilon^{ }_q)]
  [1 + n^{ }_{\sigma_5}(E^{ }_r)]
 - 
  [1 + n^{ }_{\sigma_4}(\epsilon^{ }_p)]
  n^{ }_{\sigma_2}(\epsilon^{ }_q) 
  n^{ }_{\sigma_5}(E^{ }_r)
  }{(\mathcal{K+P})^2 (\mathcal{K-Q})^2 } \biggr\} 
 \hspace*{0.29cm} \mbox{(r2)} \nn 
 & + & 
 \int_{\vec{p,q,r}}
 \frac{(2\pi)^4 \delta^{(4)}(-\mathcal{K}+\mathcal{P-Q+R})} 
 {8 \epsilon^{ }_p \epsilon^{ }_q E^{ }_r}
 \nn & & \quad \times \, \mathbbm{P} \, \biggl\{ 
 \frac{
  [1 + n^{ }_{\sigma_4}(\epsilon^{ }_p)]
  n^{ }_{\sigma_2}(\epsilon^{ }_q) 
  [1 + n^{ }_{\sigma_5}(E^{ }_r)] - 
  n^{ }_{\sigma_4}(\epsilon^{ }_p) 
  [1 + n^{ }_{\sigma_2}(\epsilon^{ }_q)]
  n^{ }_{\sigma_5}(E^{ }_r)
  }{(\mathcal{K-P})^2 (\mathcal{K+Q})^2 } \biggr\}
 \hspace*{0.29cm} \mbox{(r3)} \nn 
 & + & 
 \int_{\vec{p,q,r}}
 \frac{(2\pi)^4 \delta^{(4)}(-\mathcal{K}+\mathcal{P+Q-R})} 
 {8 \epsilon^{ }_p \epsilon^{ }_q E^{ }_r}
 \nn & & \quad \times \, \mathbbm{P} \, \biggl\{
 \frac{
  [1 + n^{ }_{\sigma_4}(\epsilon^{ }_p)]
  [1 + n^{ }_{\sigma_2}(\epsilon^{ }_q)]
  n^{ }_{\sigma_5}(E^{ }_r)
 - 
  n^{ }_{\sigma_4}(\epsilon^{ }_p) 
  n^{ }_{\sigma_2}(\epsilon^{ }_q) 
  [1 + n^{ }_{\sigma_5}(E^{ }_r)]
  }{(\mathcal{K-P})^2 (\mathcal{K-Q})^2 } \biggr\}
 \hspace*{0.30cm} \mbox{(r4)} \nn 
 & + & 
 \int_{\vec{p}}
 \frac{2\pi \delta(-\ko  + \epsilon^{ }_{pk} + \epsilon^{ }_p )}
 {4\epsilon^{ }_p \epsilon^{ }_{pk}}
 \Bigl[ 1 + n^{ }_{\sigma_4}(\epsilon^{ }_{pk}) 
 + n^{ }_{\sigma_1}(\epsilon^{ }_{p}) \Bigr] 
 \nn & & \quad \times \, 
 \int_{\vec{q}}
 \mathbbm{P}\biggl\{
   \frac{\fr12 + n^{ }_{\sigma_2} (\epsilon^{ }_q) }{2\epsilon^{ }_q} 
   \left. \frac{1}
   {[(\mathcal{Q-P})^2-\lambda^2] (\mathcal{Q-K})^2} 
    \right|^{ }_{q_0 = \pm \epsilon^{ }_q}
  \hspace*{2.92cm} \mbox{(v1)}
  \nn & & \hspace*{1.5cm} + \, 
   \frac{\fr12 + n^{ }_{\sigma_3} (\epsilon^{ }_{qk}) }{2\epsilon^{ }_{qk}} 
   \left. \frac{1}
   {[(\mathcal{Q-P})^2-\lambda^2] \mathcal{Q}^2} 
    \right|^{ }_{q_0 = \ko \pm \epsilon^{ }_{qk} }
  \hspace*{3.37cm} \mbox{(v2)}
  \nn & & \hspace*{1.5cm} + \, 
   \frac{\fr12 + n^{ }_{\sigma_5} (E^{ }_{qp}) }{2E^{ }_{qp}} 
   \left. \frac{1}
   {\mathcal{Q}^2 (\mathcal{Q-K})^2 } 
    \right|^{ }_{q_0 = p_0 \pm E^{ }_{qp} }
 \biggr\}^{ }_
  {p_0 = \epsilon^{ }_p,\, \ko = \epsilon^{ }_p + \epsilon^{ }_{pk} }
 \hspace*{1.61cm} \mbox{(v3)} \nn 
 & + & 
 \int_{\vec{p}}
 \frac{2\pi \delta(-\ko - \epsilon^{ }_{pk} + \epsilon^{ }_{p} )}
 {4\epsilon^{ }_p \epsilon^{ }_{pk}}
 \Bigl[ n^{ }_{\sigma_4}(\epsilon^{ }_{pk}) 
 - n^{ }_{\sigma_1}(\epsilon^{ }_{p}) \Bigr] 
 \nn & & \quad \times \, 
 \int_{\vec{q}}
 \mathbbm{P}\biggl\{
   \frac{\fr12 + n^{ }_{\sigma_2} (\epsilon^{ }_q) }{2\epsilon^{ }_q} 
   \left. \frac{1}
   {[(\mathcal{Q-P})^2-\lambda^2] (\mathcal{Q-K})^2} 
    \right|^{ }_{q_0 = \pm \epsilon^{ }_q}
  \nn & & \hspace*{1.5cm} + \, 
   \frac{\fr12 + n^{ }_{\sigma_3} (\epsilon^{ }_{qk}) }{2\epsilon^{ }_{qk}} 
   \left. \frac{1}
   {[(\mathcal{Q-P})^2-\lambda^2] \mathcal{Q}^2} 
    \right|^{ }_{q_0 = \ko \pm \epsilon^{ }_{qk} }
  \nn & & \hspace*{1.5cm} + \, 
   \frac{\fr12 + n^{ }_{\sigma_5} (E^{ }_{qp}) }{2E^{ }_{qp}} 
   \left. \frac{1}
   {\mathcal{Q}^2 (\mathcal{Q-K})^2 } 
    \right|^{ }_{q_0 = p_0 \pm E^{ }_{qp} }
 \biggr\}^{ }_
  {p_0 = \epsilon^{ }_p,\, \ko = \epsilon^{ }_p - \epsilon^{ }_{pk} }
 \hspace*{1.2cm} \mbox{ }
 \nn[3mm] 
 && \; + \, (\ko \to -\ko)
 \, + \, (\sigma_2 \leftrightarrow \sigma_1, 
 \sigma_4 \leftrightarrow \sigma_3)  \;. \la{rho_Ij}
\ea
Here 
$\mathbbm{P}$ refers to principal value integration which 
renders changes of integration variables unproblematic; and 
\ba
 && 
 \mathcal{K} \equiv (\ko,\vec{k}) \;, \quad
 \mathcal{P} \equiv (\epsilon^{ }_p,\vec{p}) \;, \quad
 \mathcal{Q} \equiv (\epsilon^{ }_q,\vec{q}) \;, \quad
 \mathcal{R} \equiv (E^{ }_r,\vec{r}) 
 \;, \\ 
 && 
 \epsilon^{ }_p \equiv |\vec{p}|
 \;, \quad
 \epsilon^{ }_{pk} \equiv |\vec{p-k}|
 \;, \quad
 E^{ }_r \equiv \sqrt{r^2 + \lambda^2}
 \;, \quad
 E^{ }_{qp} \equiv \sqrt{(\vec{q-p})^2 + \lambda^2}
 \;. \hspace*{1.5cm}
\ea
Furthermore the phase space distributions are defined as 
\ba
 n^{ }_\sigma(\epsilon) & \equiv & \frac{\sigma}{e^{\epsilon/T} - \sigma}
 \;, \quad \sigma = \pm 1 \;; 
 \nn  
 \quad n^{ }_+(\epsilon) & = & \nB{}(\epsilon) 
   \; \equiv \; \frac{1}{e^{\epsilon/T} - 1} 
 \;, \quad
 \quad n^{ }_-(\epsilon) \; = \; - \nF{}(\epsilon)
  \; \equiv \;
  \frac{-1}{e^{\epsilon/T} + 1} 
 \;. \la{ni}
\ea
The notation $(...)|_{q_0 = \pm \epsilon^{ }_q}$ implies
that terms with both signs be summed together.
The channels labelled by (r1)--(r4) are referred to as 
{\em real corrections}; (v1)--(v3) as 
{\em virtual corrections}. The corresponding physical 
processes are illustrated in \fig\ref{fig:amplitudes}.

%
\begin{figure}[p]
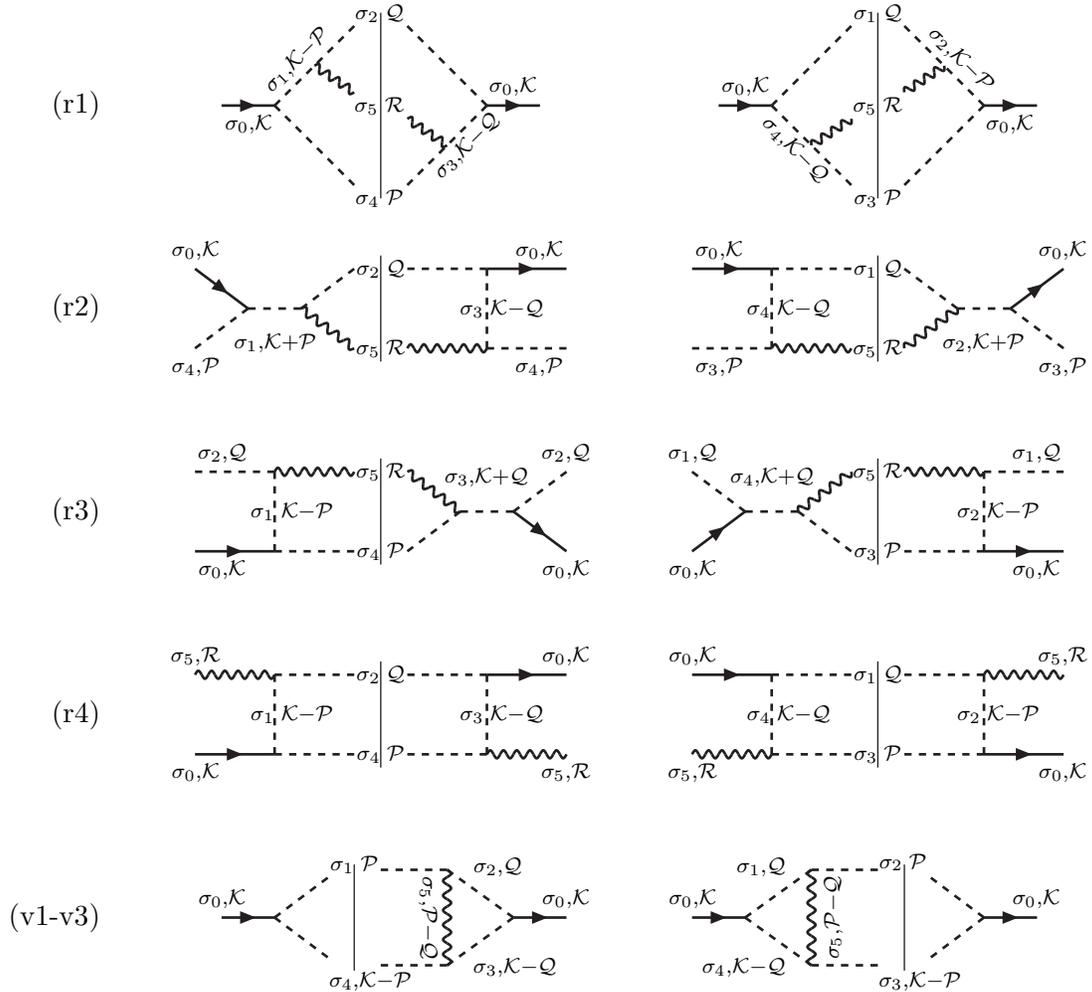


\begin{eqnarray*}
 (\mbox{r1}) && \; \ronea \qquad \roneb  \\[1mm] 
 (\mbox{r2}) && \; \rfoura \qquad \rfourb  \\[1mm] 
 (\mbox{r3}) && \; \rtwoa \qquad \rtwob  \\[1mm] 
 (\mbox{r4}) && \; \rthreea \qquad \rthreeb  \\[1mm] 
 (\mbox{v1-v3}) &&  \; \va \qquad \vb  
\end{eqnarray*}

\caption[a]{\small 
Graphical illustrations of the cut of \fig\ref{fig:Ij}, 
given explicitly in \eq\nr{rho_Ij}. Only the decay channels are shown, 
but as the phase space distributions in \eq\nr{rho_Ij} indicate, the 
inverse processes are included as well. The processes (r2), (r3), (r4) 
represent interference terms between $(u,s)$, $(t,s)$ and $(t,u)$ 
channels. For the virtual corrections the labelling (v1)--(v3) 
refers to cases in which different 
internal lines of the closed loop have a thermal weight. 
} 
\la{fig:amplitudes}
\end{figure}
%

The last line of \eq\nr{rho_Ij} adds terms with  
$\ko \to -\ko$, which can be shown to render the spectral function 
antisymmetric in this exchange. 
In the following we restrict to $\ko > 0$, and in this 
case only the terms (r1)--(r4) and (v1)--(v3) contribute. 
(The unlabelled virtual corrections below (v3)
in \eq\nr{rho_Ij} only contribute
in the space-like domain.)

%
\section{Real corrections}
\la{se:real}

Many numerical evaluations of phase space 
integrals like (r1)--(r4) can be found 
in the literature, but we are not aware of a previous
representation as a 2-dimensional integral. 
As is now demonstrated, a fairly explicit 
expression can be obtained even for $\lambda\neq 0$.

%
\subsection{Integration variables}

As is common in particle kinematics, a fruitful approach is 
to represent a complicated phase space as a convolution of simpler
ones~\cite{bk}. Taking case (r1) from \eq\nr{rho_Ij} as an example, 
we may rewrite it as  
\ba
 \rho^{ }_\rmi{r1}
 & \equiv & \int_{\vec{p,q,r}} \frac
 {(2\pi)^4 \delta(-\ko + \epsilon^{ }_p + \epsilon^{ }_q + E^{ }_r )
 \delta^{(3)}(\vec{-k+p+q+r})}
 {8\epsilon^{ }_p \epsilon^{ }_q E^{ }_r}
 \, \Phi^{ }_\rmi{r1}(\epsilon^{ }_p | \epsilon^{ }_q | E^{ }_r | \cdot )
 \nn & = &
\int_{\vec{p,q}} \frac
 {2\pi \delta(-\ko + \epsilon^{ }_{pk} + \epsilon^{ }_q + E^{ }_{qp} )}
 {8\epsilon^{ }_{pk} \epsilon^{ }_q E^{ }_{qp}}
 \, \Phi^{ }_\rmi{r1}
 (\epsilon^{ }_{pk} | \epsilon^{ }_q | E^{ }_{qp} | \cdot )
 \nn & = &
 \int_{-\infty}^{\infty}
 \! {\rm d}p_0 \, \int_{\vec{p,q}} \frac
 {2\pi \delta(-\ko + \epsilon^{ }_{pk} + p_0) 
 \delta(-p_0 + \epsilon^{ }_q + E^{ }_{qp} )}
 {8\epsilon^{ }_{pk} \epsilon^{ }_q E^{ }_{qp}}
 \, \Phi^{ }_\rmi{r1}
 (\ko - p_0 | \epsilon^{ }_q | p_0 - \epsilon^{ }_q | \cdot )
 \;, \hspace*{1cm} \la{r1}
\ea
where in the first step we substituted $\vec{p}\to \vec{k-p}$ and
integrated over $\vec{r}$; and in the second step introduced a variable
$p_0$ implementing the convolution. The function $\Phi^{ }_\rmi{r1}$ contains 
phase space distributions as well as propagators; the symbol
``$\cdot$'' stands for variables not shown explicitly. 

The key observation now is that the first three arguments 
of $\Phi^{ }_\rmi{r1}$, 
which appear in the phase space distributions, do not contain the 
variable $p$. Therefore, it is advantageous to take $p_0$ and $q$
as the outer integration variables; and $p$ as well as one 
azimuthal angle that is not fixed by the $\delta$-constraints 
as the inner ones. The inner integrations contain no phase space
distributions and, as we will see, can be carried out explicitly. 
They also turn out to capture collinear 
phase space singularities in a manageable form. 

The only challenge with this strategy is that it is tedious to 
work out the limits of the $p$-integration in the plane $(p_0,q)$.
Nevertheless, with some work, this challenge is surmountable. 
Employing the labelling shown in \fig\ref{fig:ranges}, the 
ranges for the case (r1) are [$k_\pm \equiv (\ko \pm k)/2$]:
\ba
 (\mbox{a}): & & p \in \Bigl(q - \sqrt{(q-p_0)^2-\lambda^2},
                 q + \sqrt{(q-p_0)^2-\lambda^2}\Bigr) \;, \la{range_a} \\
 (\mbox{b}): & & p \in \Bigl(q - \sqrt{(q-p_0)^2-\lambda^2},
                 2\kp - p_0 \Bigr) \;, \\
 (\underline{\mbox{b}}): & & p \in \Bigl(2\km - p_0,
                 q + \sqrt{(q-p_0)^2-\lambda^2} \Bigr) \;, \\
 (\mbox{c}): & & p \in \Bigl(2\km - p_0,
                 2\kp - p_0 \Bigr) \;, \\
 (\tilde{\mbox{c}}): & & p \in \Bigl(p_0 - 2\km,
                 q + \sqrt{(q-p_0)^2-\lambda^2} \Bigr) \;, \\
 (\mbox{d}): & & p \in \Bigl(- q + \sqrt{(q-p_0)^2-\lambda^2},
                 2\kp - p_0 \Bigr) \;, \\
 (\mbox{e}): & & p \in \Bigl(p_0 - 2\km ,
                 2\kp - p_0 \Bigr) \;, \\
 (\underline{\mbox{e}}): & & p \in \Bigl(-q + \sqrt{(q-p_0)^2-\lambda^2},
                 q + \sqrt{(q-p_0)^2-\lambda^2}\Bigr) \;. 
 \la{range_e}
\ea 
As it turns out, the situations $\ko < 2k + \sqrt{k^2 + \lambda^2}$ 
and $\ko > 2k + \sqrt{k^2 + \lambda^2}$ need 
to be handled separately, and this leads
to the two cases (c) and ($\tilde{\rm c}$).

\begin{figure}[t]


\centerline{%
\epsfysize=7.5cm\epsfbox{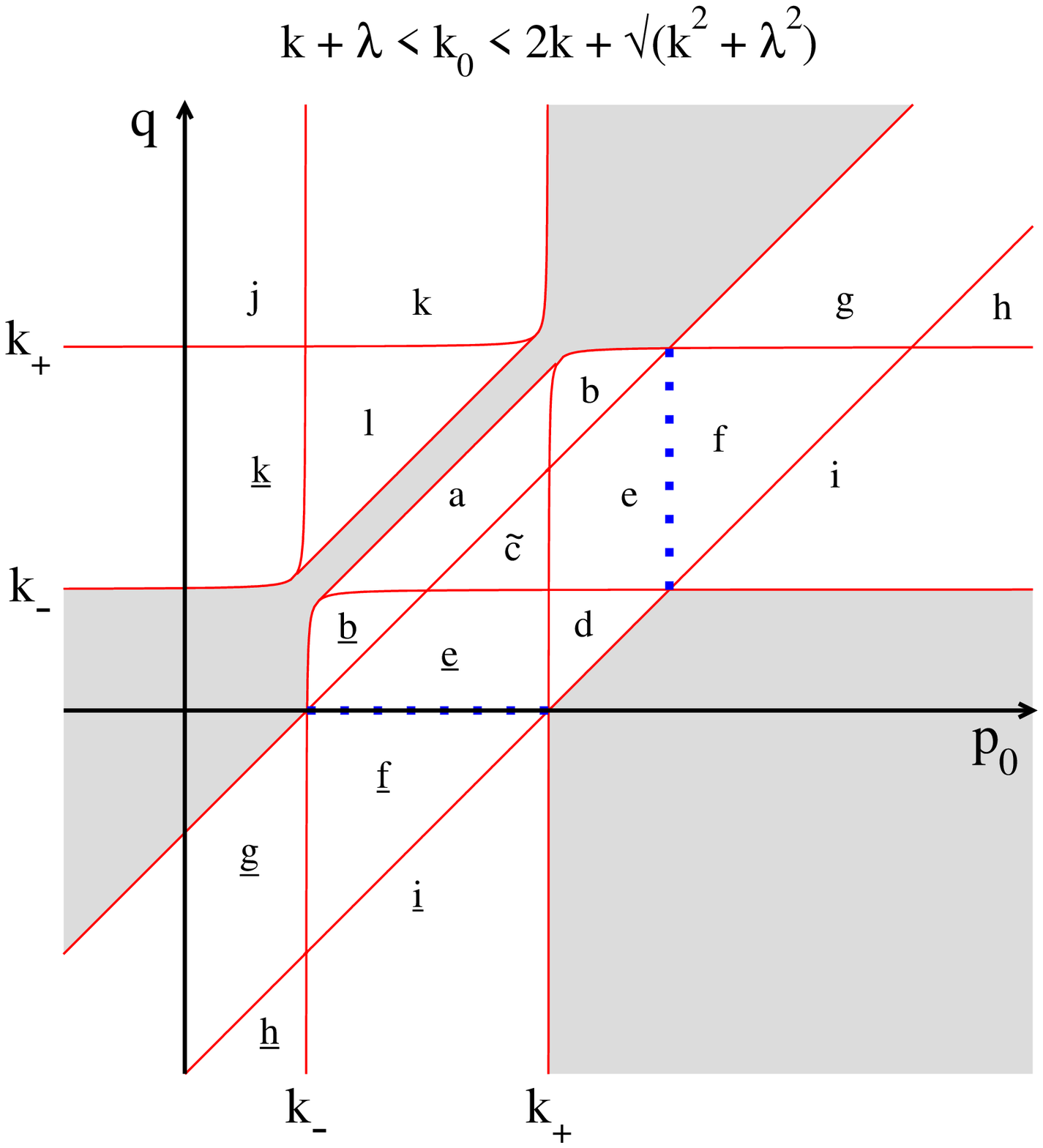}%
~~\epsfysize=7.5cm\epsfbox{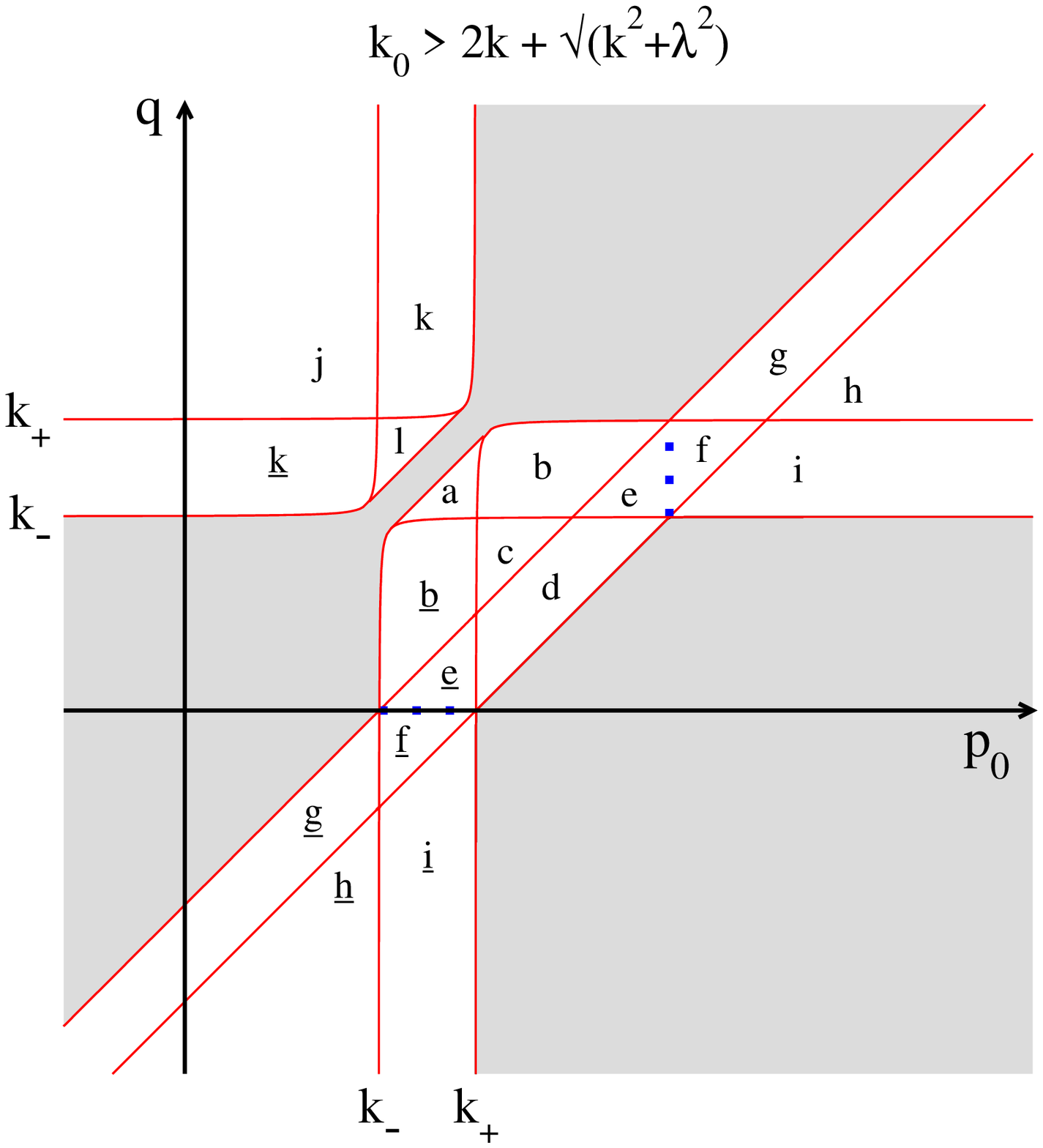}%
}

\caption[a]{\small 
 Integration ranges in the ($p_0,q$)-plane for real corrections 
 (for $\lambda = \ko/20$). The delimiting curves are 
 $
  q = \kp - \lambda^2/[4(p_0-\kp)]
 $,   
 $
  q = \km - \lambda^2/[4(p_0-\km)]
 $,   
 $
  q = p_0 - \km - \lambda^2/(4 \km)
 $, 
 $
  q = p_0 - \kp - \lambda^2/(4 \kp)
 $, 
 as well as 
 $
 q = p_0 \pm \lambda
 $.
 The various channels, separated by the dotted lines and the gap, are 
 in the middle (r1); at $p_0 > \ko$ (r2); at $q < 0$ (r3); 
 and at $q > p_0$ (r4).
 }
\la{fig:ranges}
\end{figure}

With the ranges at hand, the integrals over the angles 
\be
 \cos\chi \equiv \frac{\vec{p\cdot k}}{pk} 
 \;, \quad
 \cos\theta \equiv \frac{\vec{p\cdot q}}{pq} 
 \la{angles}
\ee
can be carried out in \eq\nr{r1}, thereby removing
the two $\delta$-functions. We are left with  
\be
 \rho^{ }_\rmi{r1} = \frac{1}{(4\pi)^3 k}
 \int^{ }_{\Omega_\rmii{r1}} 
 \!\!\! {\rm d}p_0\, {\rm d}q 
 \int_{p_\rmii{min}(p_0,q)}^{p_\rmii{max}(p_0,q)}
 \!\!\! {\rm d}p \, 
 \Bigl\langle 
   \Phi^{ }_\rmi{r1}(\ko - p_0 | q | p_0 - q | \cdot)
 \Bigr\rangle
 \;. \la{r1_2}
\ee
Here $\langle ... \rangle$ refers to an azimuthal average, 
and $\Omega^{ }_\rmi{r1}$ is composed of 
the domains (a)--(\underline{e}) of \fig\ref{fig:ranges}.

It is a nice crosscheck to set $\Phi^{ }_\rmi{r1}\to 1$ and
carry out the remaining integrals. In the absence of phase 
space distributions, the integral is Lorentz-invariant and
can alternatively be performed in a suitable frame to find a simple result: 
\be
 \left.  \rho^{ }_\rmi{r1}  \right|_{\Phi^{ }_\rmii{r1}\to 1}
 = 
 \int_{\vec{p,q,r}} \frac{(2\pi)^4 \delta^{(4)}
 \delta(-\mathcal{K+P+Q+R})}{8 \epsilon^{ }_p \epsilon^{ }_q E^{ }_r} 
 = 
 \frac{1}{(4\pi)^3}
 \biggl[ 
  \frac{\mathcal{K}^2}{4} -  
  \frac{\lambda^4}{4\mathcal{K}^2} + 
  \frac{\lambda^2}{2} \ln \frac{\lambda^2}{\mathcal{K}^2}  
 \biggr]
 \;.
\ee
This can be reproduced from \eq\nr{r1_2}, 
both for $\ko < 2 k + \sqrt{k^2+\lambda^2}$ and
$\ko > 2 k + \sqrt{k^2+\lambda^2}$.
However, in the thermal case Lorentz symmetry is not 
particularly helpful because
the plasma defines a special frame, and 
we need to make use of \eqs\nr{range_a}--\nr{range_e}.

The other channels can be handled similarly. 
For the case (r2), the shift $\vec{p}\to\vec{p-k}$
and a subsequent integration over $\vec{r}$ yields 
\ba
 \rho^{ }_\rmi{r2}
 & \equiv & \int_{\vec{p,q,r}} \frac
 {(2\pi)^4 \delta(-\ko - \epsilon^{ }_p + \epsilon^{ }_q + E^{ }_r )
 \delta^{(3)}(\vec{-k-p+q+r})}
 {8\epsilon^{ }_p \epsilon^{ }_q E^{ }_r}
 \, \Phi^{ }_\rmi{r2}(\epsilon^{ }_p | \epsilon^{ }_q | E^{ }_r | \cdot )
 \nn & = & 
 \int_{-\infty}^{\infty}
 \! {\rm d}p_0 \, \int_{\vec{p,q}} \frac
 {2\pi \delta(-\ko - \epsilon^{ }_{pk} + p_0) 
 \delta(p_0 - \epsilon^{ }_q - E^{ }_{qp} )}
 {8\epsilon^{ }_{pk} \epsilon^{ }_q E^{ }_{qp}}
 \, \Phi^{ }_\rmi{r2}
 (p_0 - \ko | \epsilon^{ }_q | p_0 - \epsilon^{ }_q | \cdot )
 \hspace*{1cm} \la{r2}
 \nn
 & = & 
\frac{1}{(4\pi)^3 k}
 \int^{ }_{\Omega_\rmii{r2}} 
 \!\!\! {\rm d}p_0\, {\rm d}q 
 \int_{p_\rmii{min}(p_0,q)}^{p_\rmii{max}(p_0,q)}
 \!\!\! {\rm d}p \, 
 \Bigl\langle 
   \Phi^{ }_\rmi{r2}(p_0 - \ko  | q | p_0 - q | \cdot)
 \Bigr\rangle
 \;. \la{r2_2}
\ea
The domain is displayed in \fig\ref{fig:ranges}, 
with the ranges
\ba
 ({\mbox{f}}): & & p \in 
                 \Bigl(2\kp - p_0 ,
                 p_0 - 2\km \Bigr)
                  \;,
 \la{range_f} \\
 ({\mbox{g}}): & & p \in 
                 \Bigl(q - \sqrt{(q-p_0)^2-\lambda^2}, 
                 p_0 - 2\km \Bigr)
                  \;, \\
 ({\mbox{h}}): & & p \in 
                 \Bigl(p_0 - 2\kp,
                 p_0 - 2\km \Bigr)
                 \;, \\
 ({\mbox{i}}): & & p \in 
                 \Bigl(- q + \sqrt{(q-p_0)^2-\lambda^2},
                 p_0 - 2\km  \Bigr) 
                 \;. 
 \la{range_i}
\ea 
For the case (r3), the shifts $\vec{p}\to\vec{k-p}$, $\vec{q}\to\vec{-q}$
and a subsequent integration over $\vec{r}$ yield 
\ba
 \rho^{ }_\rmi{r3}
 & \equiv & \int_{\vec{p,q,r}} \frac
 {(2\pi)^4 \delta(-\ko + \epsilon^{ }_p - \epsilon^{ }_q + E^{ }_r )
 \delta^{(3)}(\vec{-k+p-q+r})}
 {8\epsilon^{ }_p \epsilon^{ }_q E^{ }_r}
 \, \Phi^{ }_\rmi{r3}(\epsilon^{ }_p | \epsilon^{ }_q | E^{ }_r | \cdot )
 \nn & = &
 \int_{-\infty}^{\infty}
 \! {\rm d}p_0 \, \int_{\vec{p,q}} \frac
 {2\pi \delta(-\ko + \epsilon^{ }_{pk} + p_0) 
 \delta(p_0 + \epsilon^{ }_q - E^{ }_{qp} )}
 {8\epsilon^{ }_{pk} \epsilon^{ }_q E^{ }_{qp}}
 \, \Phi^{ }_\rmi{r3}
 (\ko - p_0 | \epsilon^{ }_q | p_0 + \epsilon^{ }_q | \cdot )
 \hspace*{1cm} \la{r3}
 \nn 
 & = & 
\frac{1}{(4\pi)^3 k}
 \int^{ }_{\Omega_\rmii{r3}} 
 \!\!\! {\rm d}p_0\, {\rm d}q 
 \int_{p_\rmii{min}(p_0,q)}^{p_\rmii{max}(p_0,q)}
 \!\!\! {\rm d}p \, 
 \Bigl\langle 
   \Phi^{ }_\rmi{r3}(\ko - p_0 | -q | p_0 - q | \cdot)
 \Bigr\rangle
 \;. \la{r3_2}
\ea
In the 2nd step we substituted formally $\epsilon^{ }_q\to -q$, 
which permits us to represent the domain 
as displayed in \fig\ref{fig:ranges}, with the ranges
\ba
 (\underline{\mbox{f}}): & & p \in \Bigl(q + \sqrt{(q-p_0)^2-\lambda^2},
                 -q + \sqrt{(q-p_0)^2-\lambda^2} \Bigr) \;,
 \la{range_uf} \\
 (\underline{\mbox{g}}): & & p \in \Bigl(2\km - p_0,
                 -q + \sqrt{(q-p_0)^2-\lambda^2} \Bigr) \;, \\
 (\underline{\mbox{h}}): & & p \in \Bigl(2\km - p_0 ,
                 2\kp - p_0 \Bigr) \;, \\
 (\underline{\mbox{i}}): & & p \in \Bigl(q + \sqrt{(q-p_0)^2-\lambda^2},
                 2\kp - p_0  \Bigr) \;. 
 \la{range_ui}
\ea 
Finally, for the case (r4), the shift $\vec{p}\to\vec{k-p}$
and a subsequent integration over $\vec{r}$ yields 
\ba
 \rho^{ }_\rmi{r4}
 & \equiv & \int_{\vec{p,q,r}} \frac
 {(2\pi)^4 \delta(-\ko + \epsilon^{ }_p + \epsilon^{ }_q - E^{ }_r )
 \delta^{(3)}(\vec{-k+p+q-r})}
 {8\epsilon^{ }_p \epsilon^{ }_q E^{ }_r}
 \, \Phi^{ }_\rmi{r4}(\epsilon^{ }_p | \epsilon^{ }_q | E^{ }_r | \cdot )
 \nn & = &
 \int_{-\infty}^{\infty}
 \! {\rm d}p_0 \, \int_{\vec{p,q}} \frac
 {2\pi \delta(-\ko + \epsilon^{ }_{pk} + p_0) 
 \delta(p_0 - \epsilon^{ }_q  + E^{ }_{qp} )}
 {8\epsilon^{ }_{pk} \epsilon^{ }_q E^{ }_{qp}}
 \, \Phi^{ }_\rmi{r4}
 (\ko - p_0 | \epsilon^{ }_q | \epsilon^{ }_q - p_0  | \cdot )
 \hspace*{1cm} \la{r4}
 \nn 
 & = & 
\frac{1}{(4\pi)^3 k}
 \int^{ }_{\Omega_\rmii{r4}} 
 \!\!\! {\rm d}p_0\, {\rm d}q 
 \int_{p_\rmii{min}(p_0,q)}^{p_\rmii{max}(p_0,q)}
 \!\!\! {\rm d}p \, 
 \Bigl\langle 
   \Phi^{ }_\rmi{r4}(\ko - p_0 | q | q - p_0 | \cdot)
 \Bigr\rangle
 \;. \la{r4_2}
\ea
The domain is displayed in \fig\ref{fig:ranges}, 
with the ranges
\ba
 ({\mbox{j}}): & & p \in 
                 \Bigl(2\km - p_0 ,
                 2\kp - p_0 \Bigr)
                  \;,
 \la{range_j} \\
 ({\mbox{k}}): & & p \in 
                 \Bigl(q - \sqrt{(q-p_0)^2-\lambda^2}, 
                 2\kp - p_0 \Bigr)
                 \;, \\
 (\underline{\mbox{k}}): & & p \in 
                 \Bigl(2\km - p_0, 
                 q + \sqrt{(q-p_0)^2-\lambda^2}  \Bigr) 
                  \;, \\
 ({\mbox{l}}): & & p \in 
                 \Bigl(q - \sqrt{(q-p_0)^2-\lambda^2},
                 q + \sqrt{(q-p_0)^2-\lambda^2} \Bigr)
                 \;. 
 \la{range_uk}
\ea 

%
\subsection{Crossing symmetry}

At zero temperature, the four channels (r1)--(r4) 
of \fig\ref{fig:amplitudes} are related by 
a crossing symmetry, and it is interesting to see how the presence
of phase space distributions in \eq\nr{rho_Ij} changes the 
situation. Inserting the arguments of $\Phi$ as displayed 
in \eqs\nr{r1_2}, \nr{r2_2}, \nr{r3_2}, \nr{r4_2}
into expressions obtained from \eq\nr{rho_Ij}, we find
\ba
 \Bigl\langle 
   \Phi^{ }_\rmi{r1}(\ko - p_0 | q | p_0 - q | \cdot)
 \Bigr\rangle
 & = & - 
 \Bigl\langle 
   \Phi^{ }_\rmi{r2}(p_0 - \ko  | q | p_0 - q | \cdot)
 \Bigr\rangle
 \nn 
 & = &  - 
 \Bigl\langle 
   \Phi^{ }_\rmi{r3}(\ko - p_0 | -q | p_0 - q | \cdot)
 \Bigr\rangle
 \nn 
 & = &  - 
 \Bigl\langle 
   \Phi^{ }_\rmi{r4}(\ko - p_0 | q | q - p_0 | \cdot)
 \Bigr\rangle
 \nn[2mm]
 & = & 
  n^{-1}_{\sigma_0}(\ko) 
 \, n^{ }_{\sigma_4} (\ko - p_0) 
 \, n^{ }_{\sigma_2} (q) 
 \, n^{ }_{\sigma_5} (p_0 - q)
 \nn[2mm] & & \times \, \mathbbm{P}\, \biggl\{ 
 \, \frac{\mathcal{K}^4}{2 (p_0^2 - p^2 )}
 \, \biggl\langle 
       \frac{1}{(\ko - q)^2 - \epsilon_{qk}^2}
    \biggr\rangle^{ }_{(\ko - p_0 | q | p_0 - q | \cdot)} \biggr\}
 \;, \hspace*{1cm}  \la{crossing} 
\ea
where on the last line the arguments ${(...|...|...|\cdot)}$
refer to $\epsilon^{ }_{pk}$, $\epsilon^{ }_q$, and $E^{ }_{qp}$, 
respectively, and we made use of the fact that the dependence
on $\epsilon^{ }_{pk}$ and $E^{ }_{qp}$ is quadratic
[cf.\ \eqs\nr{alpha}, \nr{cchi1}, \nr{ctheta1} below]
and that the sign of $q$ inside $\epsilon^{ }_{qk}$ plays no role. 
To arrive at \eq\nr{crossing} 
the conservation of fermion number 
[cf.\ \eq\nr{i_ids}], leading to 
$
 \sigma_2 \sigma_5 = \sigma_1
$ etc, 
as well as an identity following from \eq\nr{ni}, 
\be
 \sigma e^{\epsilon/T} n^{ }_\sigma (\epsilon) = - n^{ }_\sigma(-\epsilon)
 \;, 
\ee 
were assumed. The universal form in \eq\nr{crossing} 
implies that the azimuthal average $\langle ... \rangle$ 
and the subsequent integration over $p$ only need to be carried out 
for one single function. 

%
\subsection{Inner integrations}
\la{ss:inner}

We now consider the integrations still to be performed, i.e.\ 
\be
 I(p^{ }_0,q) \equiv
 \int_{p^{ }_\rmii{min}(p_0,q)}^{p^{ }_\rmii{max}(p_0,q)} \!\! {\rm d}p\,
 \mathbbm{P} \biggl\{ 
 \, \frac{\mathcal{K}^4}{2 (p_0^2 - p^2 )}
 \, \biggl\langle 
       \frac{1}{(\ko - q)^2 - \epsilon_{qk}^2}
    \biggr\rangle^{ }_{(\ko - p_0 | q | p_0 - q | \cdot)}  
 \biggr\}
 \;. \la{inner}
\ee
Let us start with the azimuthal average. 
Parametrizing\footnote{%
 Note that this parametrization can be used both for positive
 and negative $q$.
 } 
\ba
 \vec{p} & = & p\, (0,0,1) \;, \la{param1} \\ 
 \vec{k} & = & k\, (\sin\chi,0,\cos\chi) \;, \\
 \vec{q} & = & q\, (\sin\theta\cos\varphi,\sin\theta\sin\varphi,\cos\theta)\;,
 \la{param3} 
\ea
the integral over $\varphi$ is readily carried out: 
\be
 \int_{-\pi}^{\pi} \frac{{\rm d}\varphi}{2\pi} 
 \,\mathbbm{P} \biggl( \frac{1}{\alpha + \beta \cos\varphi} \biggr) = 
 \re \biggl( \frac{\sign(\alpha)}{\sqrt{\alpha^2-\beta^2}} \biggr)
 \;, \la{azimuthal}
\ee
where 
\be
 \alpha = (\ko - q)^2 - k^2 - q^2 + 2 k q \cos\chi \, \cos\theta
 \;, \quad
 \beta = 2 k q \sin \chi \, \sin \theta
 \;. \la{alpha}
\ee
Furthermore, according to \eq\nr{angles},  
the angles can be written as 
\ba
 \cos\chi  =  \frac{p^2+k^2 - \epsilon^2_{pk}}{2 p k} & = &  
 1 + \frac{(p-p_0+2 \km)(p+p_0-2\kp)}{2pk} 
 \la{cchi1} \\
 & = & 
 -1 + \frac{(p-p_0+2 \kp)(p+p_0-2\km)}{2pk} 
 \;, \la{cchi2} \\ 
 \cos\theta  =  \frac{p^2 + q^2 + \lambda^2 - E^2_{qp}}{2 qp}
 & = & 
 1 + \frac{(p-q)^2-[(q-p_0)^2-\lambda^2]}{2 p q}
 \la{ctheta1} \\
 & = & 
 - 1 + \frac{(p+q)^2-[(q-p_0)^2-\lambda^2]}{2 p q}
 \;, \la{ctheta2}
\ea
where $\epsilon^{ }_{pk}$, $E^{ }_{qp}$ were 
inserted from the arguments shown in \eq\nr{inner}.  
It can seen 
that at the boundaries of the $p$-integration, cf.\ 
\eqs\nr{range_a}--\nr{range_e}, 
\nr{range_f}--\nr{range_i},   
\nr{range_uf}--\nr{range_ui},   
\nr{range_j}--\nr{range_uk}, 
one of the cosines evaluates to $\pm 1$, and correspondingly
one of the sines vanishes. Therefore,
at the boundaries the function $\beta$ given in \eq\nr{alpha} vanishes; 
this observation will turn out to be useful in a moment. 

Inspecting the expressions it is now possible to realize that, 
in general,  the dependence of $\alpha^2-\beta^2$ appearing
in \eq\nr{azimuthal} on $p$ is of the form 
\be
  \alpha^2 - \beta^2 = a \, p^2 + b + \frac{c}{p^2} 
  \;. \la{prms}
\ee 
Here $a = (\ko - q)^2$ and the other coefficients are more
complicated. Remarkably, this functional form implies that 
the integral defined in \eq\nr{inner}  
can be carried out:\footnote{%
  The function $\alpha$ can be positive or negative but not 
  change its sign within the ranges considered. 
  } 
\ba
 I(p_0,q) & = & \frac{\mathcal{K}^4}{4}
 \int_{p^{ }_\rmii{min}}^{p^{ }_\rmii{max}} 
 \frac{2 p\, {\rm d}p}{p_0^2 - p^2} 
 \re\biggl\{ \frac{\sign(\alpha)}{\sqrt{a p^4 + b p^2 + c}} \biggr\}
 \nn 
 & = & 
 \frac{\mathcal{K}^4\sign(\alpha)}{4\sqrt{a p_0^4 + b p_0^2 + c}}
 \biggl\{ 
   \ln \biggl| \frac{p_0^2 - p_\rmii{min}^2}{p_0^2 - p_\rmii{max}^2} \biggr|
 \nn & & 
 + \, \ln \biggl|
          \frac{(\sqrt{a p_0^4 + b p_0^2 + c} +
                 \sqrt{a p_\rmii{max}^4 + b p_\rmii{max}^2 + c}  )^2
                  - a (p_0^2 - p_\rmii{max}^2)^2}
               {(\sqrt{a p_0^4 + b p_0^2 + c} +
                 \sqrt{a p_\rmii{min}^4 + b p_\rmii{min}^2 + c}  )^2
                  - a (p_0^2 - p_\rmii{min}^2)^2} 
       \biggr|
 \biggr\}
 \;.
\ea
Furthermore, both square roots can be simplified:
the prefactor contains the function 
\ba
 \mathcal{F}(p_0,q) & \equiv &  \sqrt{ a p_0^4 + b p_0^2 + c } 
 \nn & = & 
 \sqrt{(q-p_0)^2\mathcal{K}^4 - \lambda^2 
 [\mathcal{K}^2 + 4 q p_0 - 2 \ko (q+p_0)]\mathcal{K}^2 + \lambda^4 k^2}
 \;, \la{calF}
\ea
whereas, as already mentioned in connection 
with \eqs\nr{cchi1}--\nr{ctheta2}, the function $\beta$ vanishes
for $p = p^{ }_\rmi{min}$ and $p = p^{ }_\rmi{max}$. Therefore, 
according to \eq\nr{prms}, 
\be 
 \sqrt{a p_\rmi{max}^4 + b p_\rmi{max}^2 + c} = 
 p_\rmi{max} |\alpha ( p^{ }_\rmi{max} ) |
 \;,
\ee
and correspondingly for $p^{ }_\rmi{min}$, where $\alpha$
is the function from \eq\nr{alpha}. 

To summarize, all spectral functions corresponding to 
real corrections have  
2-dimensional integral representations: 
\ba
 \rho^{ }_\rmi{r1} & = &  
 \frac{\pi \mathcal{K}^4   n^{-1}_{\sigma_0}(\ko) }{(4\pi)^4 k}
 \int^{ }_{\Omega_\rmii{r1}} 
 \!\!\! {\rm d}p_0\, {\rm d}q 
 \, 
 \, n^{ }_{\sigma_4} (\ko - p_0) 
 \, n^{ }_{\sigma_2} (q) 
 \, n^{ }_{\sigma_5} (p_0 - q)
 \nn & & \hspace*{-1cm} \times \, 
 \frac{\sign(\alpha)}{\mathcal{F}(p_0,q)}
 \biggl\{ 
   \ln \biggl| \frac{p_0^2 - p_\rmii{min}^2}{p_0^2 - p_\rmii{max}^2} \biggr|
 + \, \ln \biggl|
          \frac{[ \mathcal{F}(p_0,q) 
                 + | p_\rmii{max} \alpha |  ]^2
                  - (q-\ko)^2 (p_0^2 - p_\rmii{max}^2)^2}
               {[ \mathcal{F}(p_0,q) 
                 + | p_\rmii{min} \alpha | ]^2
                  - (q-\ko)^2 (p_0^2 - p_\rmii{min}^2)^2} 
       \biggr|
 \biggr\}
 \;, \nn \la{r1_3}
\ea
and correspondingly for 
$
 \rho^{ }_\rmi{r2}
$, 
$
 \rho^{ }_\rmi{r3}
$ and
$
 \rho^{ }_\rmi{r4}
$.
It can be recalled from \eq\nr{crossing} that the 
other channels come with an overall minus sign, and we also 
find that $\sign(\alpha) = -1$ for the channels (r2) and (r4).
The integration range 
$\Omega^{ }_\rmi{r1} + \Omega^{ }_\rmi{r2} + \Omega^{ }_\rmi{r3}
+ \Omega^{ }_\rmi{r4}$ is as given in \fig\ref{fig:ranges}.

Although well suited for numerical handling, 
the integral representation
in \eq\nr{r1_3} remains fairly complicated in practice as long as
$\lambda\neq 0$. For $\lambda\to 0$, the expression simplifies but 
is also logarithmically divergent. However, when we sum the result 
together with virtual corrections, to which we now turn, the 
divergences cancel; the resulting expressions are presented
in \se\ref{se:sum}.

%
\section{Virtual corrections}
\la{se:virtual}

For $\ko > k > 0$, the virtual corrections are 
contained within the terms denoted by (v1)--(v3) in \eq\nr{rho_Ij}, 
as well as in their reflections 
$
 (\sigma_2 \leftrightarrow \sigma_1, \sigma_4 \leftrightarrow \sigma_3)
$.
Each of the terms is factorized into two structures. 
In the first one 
the angular integration is immediately doable, 
and thereby we can simplify the radial integration measure into 
\be
 \int_{\vec{p}}
 \frac{2\pi \delta(-\ko + \epsilon^{ }_p + \epsilon^{ }_{pk} )}
 {4\epsilon^{ }_p \epsilon^{ }_{pk}}
 \Bigl[ 1 + n^{ }_{\sigma_4}(\epsilon^{ }_{pk}) 
 + n^{ }_{\sigma_1}(\epsilon^{ }_{p}) \Bigr] 
 = \frac{n^{-1}_{\sigma_0}(\ko)}{8\pi k}
 \int_{\km}^{\kp} \! {\rm d}p \, 
 n^{ }_{\sigma_4} (\ko - p) n^{ }_{\sigma_1}(p)
 \;. \la{fz_p1}
\ee
As far as the $\vec{q}$-integrals 
are concerned, their angular parts 
can be carried out with the help of a Feynman parameter, 
$s \in (0,1)$.\footnote{%
 The vacuum part, which is represented by the factors $\fr12$
 in \eq\nr{rho_Ij}, could be integrated explicitly by making use
 of Lorentz invariance 
 (the result is shown in \eq\nr{triangle} below), however  
 for us it is convenient to handle it together with 
 the thermal contributions. 
 For a number of other master spectral  
 functions the vacuum part is divergent at large $|\vec{q}|$, but it turns 
 out that even in those cases it is convenient to handle it together
 with the thermal contributions for moderate $|\vec{q}|\lsim \ko$; 
 only the asymptotics at $|\vec{q}|\gg \ko$ needs to be handled separately
 with a proper ultraviolet regularization. 
 }
Considering first the case (v1), we note that 
\be
   s [(\mathcal{Q-P})^2-\lambda^2] + (1-s) (\mathcal{Q-K})^2
  =  m_s^2 
    + 2 \vec{q} \cdot \vec{e}_s 
  \;, \la{rpf_2}
\ee
where 
$
 m_s^2 = (1-s) \mathcal{K}^2
    - s \lambda^2 
  - 2 q_0 [s p + (1-s) \ko]
$ 
and 
$\vec{e}_s \equiv s\vec{p} + (1-s) \vec{k}$. 
Denoting by $z$
the angle between $\vec{q}$ and $\vec{e}_s$, we trivially get 
\be
 \int_{-1}^{+1} \! \frac{{\rm d}z}{2}
 \frac{1}{(m_s^2 + 2 q e_s z)^2} = \frac{1}{m_s^4 - 4 q^2 e_s^2}
 \;. \la{rpf_1}
\ee  
Furthermore, by making use of the constraint 
$\ko = p + \epsilon^{ }_{pk}$ implied by \eq\nr{fz_p1}, 
the length of $\vec{e}_s$ is given by 
$
 e_s^2 = [s p + (1-s) \ko]^2 - (1-s)\mathcal{K}^2
$.
The denominator in \eq\nr{rpf_1} is thus a 2nd order polynomial in $s$, 
and the integral over $s$ is also doable: 
\be
 \int_0^1  \frac{{\rm d}s}{\mu s^2 - 2 \nu s + \rho}
 = \frac{1}{2\sqrt{\nu^2 - \mu \rho}}
   \ln \frac{\nu-\rho - \sqrt{\nu^2 - \mu \rho}}
            {\nu-\rho + \sqrt{\nu^2 - \mu \rho}}
 \;.
\ee 
As a final step, the terms with $q_0 = \pm \epsilon^{ }_q$ 
can formally be combined by making use of 
\be
 \fr12 + n^{}_{\sigma_2}(-q) = - \Bigl[ \fr12 + n^{ }_{\sigma_2}(q) \Bigr]
 \;. \la{n_id_x}
\ee
Thereby the term (v1) can be cast in the form
\ba
 \rho^{ }_\rmi{v1} & = &  \frac{\pi \mathcal{K}^4 n^{-1}_{\sigma_0}(\ko)}
 {(4\pi)^4 k}
 \int_{\km}^{\kp} \!{\rm d}p 
 \, n^{ }_{\sigma_4}(\ko - p) n^{ }_{\sigma_1}(p) 
 \int_{-\infty}^{\infty} \! {\rm d}q \, 
 \Bigl[ \fr12 + n^{ }_{\sigma_2}(q) \Bigr] 
 \nn & & \times \, 
 \frac{1}{\mathcal{F}(p,q)} \ln 
 \biggl| \frac{2 \mathcal{K}^2 q (q-p)
  - \lambda^2(\mathcal{K}^2 - 2 \ko q) + 2 q \mathcal{F}(p,q)}
              {2 \mathcal{K}^2 q (q-p)
  - \lambda^2(\mathcal{K}^2 - 2 \ko q) - 2 q \mathcal{F}(p,q)} \biggr|
 \;, \la{v1}
\ea
where $\mathcal{F}$ is the same function as appeared
in the real corrections, given by \eq\nr{calF}.
The argument of the square root in $\mathcal{F}$ is always positive, so that 
the integrand is well-defined;\footnote{%
 It may be noted, however, that the integrand has a non-trivial 
 structure at small $q$, with the argument of the logarithm
 having a zero at $q = - \lambda^2/(4p)$. This (integrable) singularity 
 cancels against a corresponding one from real 
 corrections, related to structure near the boundaries between the regimes
 (r1) and (r3) in \fig\ref{fig:ranges}, cf.\ footnote~\ref{subtle}.
 } 
however, the vacuum part
of the $q$-integral in \eq\nr{v1} is divergent on its own, and
the integral should only be carried out for 
the {\em sum} of the three terms (v1)--(v3).
 
The second structure (v2) can be reduced to the first one 
by substituting variables as $\vec{p}\to \vec{k-p}$, 
$\vec{q}\to\vec{k-q}$ in \eq\nr{rho_Ij}. Effectively, this corresponds to  
an interchange $\sigma_4 \leftrightarrow \sigma_1$ in \eq\nr{v1}.
However, we can subsequently also substitute $p\to \ko -p$, 
$q\to \ko - q$, and given that $\mathcal{F}(p,q)$ of \eq\nr{calF} 
is invariant in this transformation, the result reads
\ba
 \rho^{ }_\rmi{v2} & = &  \frac{\pi \mathcal{K}^4  n^{-1}_{\sigma_0}(\ko)}
 {(4\pi)^4 k} 
 \int_{\km}^{\kp} \!{\rm d}p  
 \, n^{ }_{\sigma_4}(\ko - p) n^{ }_{\sigma_1}(p) 
 \int_{-\infty}^{\infty} \! {\rm d}q 
 \Bigl[ \fr12 + n^{ }_{\sigma_3}(\ko - q) \Bigr] 
 \nn & & \hspace*{-1cm} \times \, 
 \frac{1}{\mathcal{F}(p,q)} \ln 
 \biggl| \frac{2 \mathcal{K}^2 (q -\ko) (q-p)
  - \lambda^2[\mathcal{K}^2 + 2 \ko (q - \ko)]
   + 2 (\ko - q) \mathcal{F}(p,q)}
              {2 \mathcal{K}^2 (q - \ko) (q-p)
  - \lambda^2[\mathcal{K}^2 + 2 \ko (q - \ko)]
 - 2 (\ko - q) \mathcal{F}(p,q)} \biggr|
 \;. \la{v2}
\ea

For the third term (v3), we can substitute $\vec{q}\to \vec{p-q}$
in \eq\nr{rho_Ij}.
Eq.~\nr{rpf_2} gets replaced with 
\be
   s (\mathcal{P-Q-K})^2 + (1-s) (\mathcal{P-Q})^2
  =  m_s^2 
    + 2 \vec{q} \cdot \vec{e}_s 
  \;, \la{rpf_3}
\ee
where now
$
 m_s^2 = \lambda^2 - 2 q_0 (p - s \ko)
$, 
$\vec{e}_s \equiv \vec{p}  -s \vec{k}$, 
and $q_0 = \pm E^{ }_q$. 
The subsequent steps go as before, noting that 
$e_s^2 = p^2 + s(\mathcal{K}^2 - 2 p \ko) + s^2 k^2$.
Afterwards, it is convenient to again return to 
the ``original'' variables; this can be implemented
by first taking $E_q$ as an integration variable instead of $q$, 
and then introducing a ``new'' $q$ as $E_q = |p-q|$. 
In this way the result can be cast in a form 
reminiscent of \eqs\nr{v1}, \nr{v2}: 
\ba
 \rho^{ }_\rmi{v3} & = &  \frac{\pi \mathcal{K}^4  n^{-1}_{\sigma_0}(\ko)}
 {(4\pi)^4 k} 
 \int_{\km}^{\kp} \!{\rm d}p 
 \, n^{ }_{\sigma_4}(\ko - p) n^{ }_{\sigma_1}(p) 
 \biggl[ \int_{-\infty}^{p-\lambda}
  + \int_{p+\lambda}^{\infty} \biggr]  {\rm d}q \, 
 \Bigl| \fr12 + n^{ }_{\sigma_5}(p - q) \Bigr| 
 \nn & & \hspace*{-1cm} \times \, 
 \frac{1}{\mathcal{F}(p,q)} \ln 
 \biggl| \frac{\mathcal{K}^2 (q-p)^2
  - \lambda^2[\mathcal{K}^2 +2 q p - \ko (q+p)] - \frac{\lambda^4}{2} 
 - \sqrt{(q-p)^2 - \lambda^2} \mathcal{F}(p,q)}
 {\mathcal{K}^2 (q-p)^2
  - \lambda^2[\mathcal{K}^2 +2 q p - \ko (q+p)] - \frac{\lambda^4}{2} 
 + \sqrt{(q-p)^2 - \lambda^2} \mathcal{F}(p,q)}
   \biggr|
 \;. 
 \nn \la{v3}
\ea

The expressions in \eqs\nr{v1}, \nr{v2}, \nr{v3} 
contain no expansion with respect to $\lambda$. 
We have crosschecked numerically in the small-$\lambda$
regime that picking up the vacuum factors 
(the ``$\fr12$''s if the $n$'s are expressed with positive arguments), 
their sum is integrable and reproduces within the achieved 
resolution a known Lorentz-invariant integral: 
\ba
  && \hspace*{-1cm} \int_Q \mathbbm{P} 
  \biggl\{ 
    \frac{1}{[(Q-P)^2 + \lambda^2 ] Q^2 (Q-K)^2}
  \biggr\}_{P^2 = 0,\, K^2 = -\mathcal{K}^2, \,  
  K \cdot P = - \frac{\mathcal{K}^2}{2}} 
 \nn
  &  = &  
 \frac{1}{(4\pi)^2 \mathcal{K}^2}
 \biggl[ 
   \frac{\pi^2}{6} - 
  \fr12 \ln^2 \biggl(\frac{\lambda^2}{\mathcal{K}^2 + \lambda^2} \biggr)
 - \Li \biggl(\frac{\lambda^2}{\mathcal{K}^2 + \lambda^2} \biggr)
 \biggr]
 \;. \la{triangle}
\ea

%
\section{Cancellation of divergences}
\la{se:sum}

The results in \ses\ref{se:real} and \ref{se:virtual} are both divergent if
we attempt to send $\lambda\to 0$ (cf.\ \eq\nr{triangle} for the vacuum part). 
Their sum, however, remains finite as we now demonstrate. 

The expressions obtained have two kinds of ``singularities''. 
In terms of \fig\ref{fig:ranges}, with variables suitably renamed
for virtual corrections, the integrands have a non-trivial 
structure around the boundaries separating the different channels, 
i.e.\ $q = p_0$, $q=0$, and $p_0 = \ko$. In addition, away from the 
boundaries, the integrands in general diverge as $\lambda\to 0$.

The strategy we adopt is to stay away from the boundaries, for 
instance by setting a band of width $\delta$ around them, and 
taking the limit $\lambda\to 0$ within the domains.\footnote{%
  Actually this requires a somewhat more careful justification, given that
  according to \eq\nr{crossing} there is a phase space distribution
  at each boundary which would diverge if it were bosonic. It turns 
  out that if the integrand of \eq\nr{r1_3} 
  is evaluated within the domains adjacent to 
  the boundaries and we make the substitution in 
  \eq\nr{subst}, then it cancels {\em exactly} against an integrand 
  in one of the 
  virtual corrections, namely that containing the same potentially
  divergent phase space distribution. In other words, the 2nd row  
  of \eq\nr{r1_3} evaluated 
  within the domains (a) and (l)
  of \fig\ref{fig:ranges} cancels against the 2nd row 
  of \eq\nr{v3}; 
  within (e) and (f) against that of \eq\nr{v2} once 
  the latter is reflected in $p\leftrightarrow q$; 
  and within ($\underline{\mbox{e}}$) and
  ($\underline{\mbox{f}}$) against that of \eq\nr{v1}. Approximate
  forms of these cancellations can be seen by contrasting
  \eqs\nr{coeff_first} and \nr{v3_sim}; 
  \nr{e} and \nr{v2_sim} [with $p\leftrightarrow q$]; 
  as well as \nr{under_e} and \nr{v1_sim}.  \la{subtle}
  }
We then 
verify the cancellation of the corresponding divergences 
within the domains, and that the resulting integrand
remains integrable even after ultimately setting  $\delta\to 0$. 

Turning first to the real corrections, \eq\nr{r1_3}, we simplify
the notation from now on by renaming 
\be
 p_0 \rightarrow p 
 \;. \la{subst}
\ee
Then the following expressions are obtained for the integrand
of \eq\nr{r1_3} {\em inside} the domains of \fig\ref{fig:ranges}:
\ba
 \frac{\sign(\alpha)}{\mathcal{F}(p,q)} \biggl\{ ... \biggr\} 
 \; \stackrel{\lambda\to 0}{\rightarrow} \; \quad
 \mbox{(a)} = - \mbox{(l)}:&& 
 \frac{1}{(p-q) \mathcal{K}^2} 
 \ln\biggl|
   \frac{\mathcal{K}^4 (p-q)^4}{\lambda^4 p q (p-\ko)(q-\ko)} 
 \biggr|
 \;, \la{coeff_first} \\
 \mbox{(b)}:&& 
 \frac{1}{(p-q) \mathcal{K}^2} 
 \ln\biggl|
   \frac{\mathcal{K}^2 (p-q)^2 (q-\kp)}{\lambda^2 q(q-\ko)(p-\kp)} 
 \biggr|
 \;, \\
 (\underline{\mbox{b}}):&& 
 \frac{1}{(p-q) \mathcal{K}^2} 
 \ln\biggl|
   \frac{\mathcal{K}^2 (p-q)^2 (p-\km)}{\lambda^2 p (p-\ko)(q-\km)} 
 \biggr|
 \;, \\
 \mbox{(c)} = \mbox{(h)} = (\underline{\mbox{h}}) = \mbox{(j)}:&& 
 \frac{1}{(p-q) \mathcal{K}^2} 
 \ln\biggl|
   \frac{(p-\km) (q-\kp)}{(p-\kp)(q-\km)} 
 \biggr|
 \;, \\
 (\tilde{\mbox{c}}):&&
 \frac{1}{(p-q) \mathcal{K}^2} 
 \ln\biggl|
   \frac{\mathcal{K}^4 (p-q)^4}
   {\lambda^4 p (q-\ko) (p-\km)(q-\kp)} 
 \biggr|
 \;, \hspace*{0.5cm} \\
 \mbox{(d)}:&& 
 \frac{1}{(p-q) \mathcal{K}^2} 
 \ln\biggl|
   \frac{q (p-\ko)}{(p-\kp)(q-\km)} 
 \biggr|
 \;, \\
 \mbox{(e)} = -\mbox{(f)}:&& 
 \frac{1}{(p-q) \mathcal{K}^2} 
 \ln\biggl|
   \frac{\mathcal{K}^2 (p-q)^2 (p-\ko)}{\lambda^2 (q-\ko)(p-\km)(p-\kp)} 
 \biggr|
 \;, \la{e} \\
 (\underline{\mbox{e}}) = -(\underline{\mbox{f}}):&& 
 \frac{1}{(p-q) \mathcal{K}^2} 
 \ln\biggl|
   \frac{\mathcal{K}^2 (p-q)^2 q }{\lambda^2 p (q-\km)(q-\kp)} 
 \biggr|
 \;, \la{under_e} \\
 -\mbox{(g)}  = -(\underline{\mbox{g}}):&& 
 \frac{1}{(p-q) \mathcal{K}^2} 
 \ln\biggl|
   \frac{q (p-\ko)}{(p-\km)(q-\kp)} 
 \biggr|
 \;, \\
 -\mbox{(i)} = -(\underline{\mbox{k}}):&& 
 \frac{1}{(p-q) \mathcal{K}^2} 
 \ln\biggl|
   \frac{\mathcal{K}^2 (p-q)^2 (q-\km)}{\lambda^2 q(q-\ko)(p-\km)} 
 \biggr|
 \;, \\
 -(\underline{\mbox{i}}) = -(\mbox{k}) :&& 
 \frac{1}{(p-q) \mathcal{K}^2} 
 \ln\biggl|
   \frac{\mathcal{K}^2 (p-q)^2 (p-\kp)}{\lambda^2 p (p-\ko)(q-\kp)} 
 \biggr|
 \;.  \la{coeff_last}
\ea
These are multiplied by phase space distributions as indicated 
by \eq\nr{crossing}, and in addition the symmetrization 
$(\sigma_2 \leftrightarrow \sigma_1, \sigma_4 \leftrightarrow \sigma_3)$
from the last line of \eq\nr{rho_Ij} needs to be included. 

\begin{figure}[t]


\centerline{%
\epsfysize=7.5cm\epsfbox{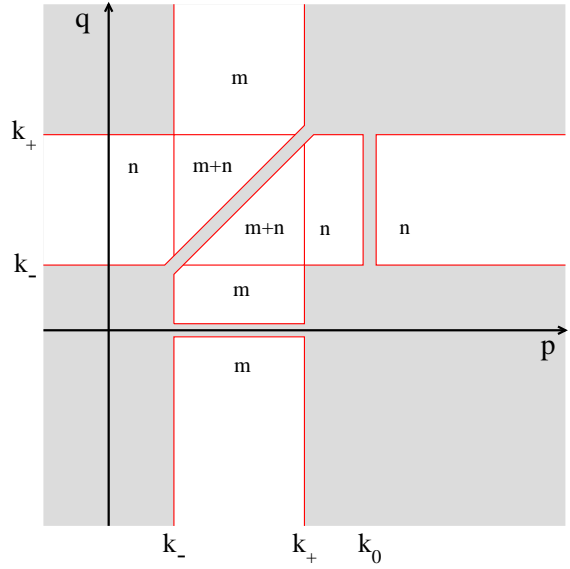}%
}

\caption[a]{\small 
 Integration ranges in the ($p,q$)-plane for 
 virtual corrections. The delimiting curves are 
 $
  p = \km 
 $,   
 $
  p = \kp 
 $,   
 $
  q = \km 
 $, 
 and 
 $
  q = \kp 
 $.
 Narrow 
 bands have been cut off around the lines
 $
 q = p
 $, 
 $
  q = 0
 $ 
 and 
 $
 p= \ko
 $. 
 }
\la{fig:virtual}
\end{figure}

Similar expressions are obtained for the virtual corrections, 
\eqs\nr{v1}, \nr{v2}, \nr{v3}: 
\ba
 \frac{1}{\mathcal{F}(p,q)} \ln\biggl| ... \biggr| 
  \; \stackrel{\lambda\to 0}{\rightarrow} \;  \quad
 (\mbox{v1}^{ }):&& 
 \frac{1}{(p-q) \mathcal{K}^2} 
 \ln\biggl|
   \frac{\lambda^2 p (q-\km)(q-\kp)} {\mathcal{K}^2 q (p-q)^2}
 \biggr|
 \;, \la{v1_sim} \\
  (\mbox{v2}^{ }):&& 
 \frac{-1}{(p-q) \mathcal{K}^2} 
 \ln\biggl|
   \frac{\lambda^2 (p -\ko) (q-\km)(q-\kp)} {\mathcal{K}^2 (q -\ko) (p-q)^2}
 \biggr| 
 \;, \hspace*{0.5cm} \la{v2_sim} \\
  (\mbox{v3}^{ }):&& 
 \frac{1}{| p-q | \mathcal{K}^2} 
 \ln\biggl|
   \frac{\lambda^4 p q (p-\ko)(q-\ko)}{\mathcal{K}^4 (p-q)^4}
 \biggr|
 \;. \la{v3_sim}
\ea
Again the symmetrization 
$(\sigma_2 \leftrightarrow \sigma_1, \sigma_4 \leftrightarrow \sigma_3)$
needs to be included. 

In order to combine \eqs\nr{v1_sim}--\nr{v3_sim} with 
the real corrections, it is beneficial to rename
integration variables.  By making use of 
\be
 n^{ }_\sigma(-\epsilon) = -1 - n^{ }_\sigma(\epsilon) \;, \quad
 n^{ }_\sigma(\epsilon) n^{ }_\tau(\delta - \epsilon) = 
 n^{ }_{\sigma\tau}(\delta) \bigl[ 
  1 + n^{ }_\sigma(\epsilon) + n^{ }_\tau(\delta - \epsilon) 
 \bigr]
 \;, \la{n_identity}
\ee
and recalling the identities 
$\sigma_2 \sigma_5 = \sigma_1$, 
$\sigma_3 \sigma_5 = \sigma_4$, 
following from \eq\nr{i_ids}, we can write 
\ba
 \fr12 + n^{ }_{\sigma_2}(q) & = & n^{-1}_{\sigma_1}(p) 
 n^{ }_{\sigma_2}(q) n^{ }_{\sigma_5}(p-q) - \fr12 - n^{ }_{\sigma_5}(p-q)
 \;, \la{n_id_1} \\ 
 \fr12 + n^{ }_{\sigma_3}(\ko-q) & = & n^{-1}_{\sigma_4}(\ko - p) 
 n^{ }_{\sigma_3}(\ko - q) n^{ }_{\sigma_5}(q-p)
 + \fr12 + n^{ }_{\sigma_5}(p-q)
 \;. \la{n_id_2}
\ea
Furthermore, in the first term on the right-hand side of \eq\nr{n_id_2}, 
in which $n^{ }_{\sigma_3}(\ko - q)$ appears, we exchange variables as
$p\leftrightarrow q$. In the terms from  
$(\sigma_2 \leftrightarrow \sigma_1, 
 \sigma_4 \leftrightarrow \sigma_3)$, 
we can also do this in the terms involving  
$\fr12 + n^{ }_{\sigma_5}(p-q)$ for more symmetry.

The effect of these rewritings is that 
logarithms of $\lambda$ disappear from terms involving 
$\fr12 + n^{ }_{\sigma_5}(p-q)$ 
[essentially its coefficient is given by 
$-$\nr{v1_sim} + \nr{v2_sim} + \nr{v3_sim}].
Logarithms of $\lambda$ do not cancel from the coefficients
of the first terms of 
\eqs\nr{n_id_1}, \nr{n_id_2} but, combining with 
the other phase space distributions from 
\eqs\nr{v1}, \nr{v2}, are seen 
to come with the same ``weight functions'' 
as in the real corrections. 
To be explicit,  
the virtual corrections 
[$\rho^{ }_\rmi{v} \equiv 
\rho^{ }_\rmi{v1} + \rho^{ }_\rmi{v2} + \rho^{ }_\rmi{v3} +
(\sigma_2 \leftrightarrow \sigma_1, \sigma_4 \leftrightarrow \sigma_3) $]
can be represented
within the domains shown in \fig\ref{fig:virtual} as 
\ba
 \frac{(4\pi)^4 k\, \rho^{ }_\rmi{v}(\mathcal{K})}
 {\pi \mathcal{K}^4 n^{-1}_{\sigma_0}(\ko)}
 & = &  
 \int^{ }_{\Omega^{ }_\rmii{m}} \!\! {\rm d}p\,{\rm d}q \, 
  \biggl\{ 
 \Bigl[ n^{ }_{\sigma_4}(\ko -p)n^{ }_{\sigma_2}(q) 
      + n^{ }_{\sigma_3}(\ko -p)n^{ }_{\sigma_1}(q) \Bigr]
 n^{ }_{\sigma_5}(p-q) 
 \nn & & \hspace*{2cm} \times \, \frac{1}{(p-q)\mathcal{K}^2}
 \ln\biggl|
   \frac{\lambda^2 p (q-\km)(q-\kp)} {\mathcal{K}^2 q (p-q)^2}
 \biggr|
 \nn & & + \quad \, 
 n^{ }_{\sigma_4}(\ko -p)n^{ }_{\sigma_1}(p) 
 \Bigl[\fr12 +  n^{ }_{\sigma_5}(p-q)  \Bigr]
 \nn & & \hspace*{2cm} \times \, \frac{2}{(p-q)\mathcal{K}^2}
 \ln\biggl| 
   \frac{q (q-\ko)}{(q-\km)(q-\kp)}
 \biggr| \; \biggr\}
 \nn
 & + &  
 \int^{ }_{\Omega^{ }_\rmii{n}} \!\! {\rm d}p\,{\rm d}q \, 
  \biggl\{ 
 \Bigl[ n^{ }_{\sigma_4}(\ko -p)n^{ }_{\sigma_2}(q) 
      + n^{ }_{\sigma_3}(\ko -p)n^{ }_{\sigma_1}(q) \Bigr]
 n^{ }_{\sigma_5}(p-q) 
 \nn & & \hspace*{2cm} \times \, \frac{1}{(p-q)\mathcal{K}^2}
 \ln\biggl|
   \frac{\lambda^2 (q-\ko) (p-\km)(p-\kp)} {\mathcal{K}^2 (p-\ko) (p-q)^2}
 \biggr|
 \nn & & + \quad \, 
 n^{ }_{\sigma_3}(\ko -q)n^{ }_{\sigma_2}(q) 
 \Bigl[\fr12 +   n^{ }_{\sigma_5}(p-q)  \Bigr]
 \nn & & \hspace*{2cm} \times \, \frac{2}{(p-q)\mathcal{K}^2}
 \ln\biggl| 
   \frac{p (p-\ko)}{(p-\km)(p-\kp)}
 \biggr| \; \biggr\}
 \;. \la{v_all}
\ea

Combining \eq\nr{v_all}
with \eqs\nr{coeff_first}--\nr{coeff_last}, 
the latter multiplied by phase space
distributions according to \eq\nr{crossing}, and adding 
for the real corrections terms from the symmetrization 
$(\sigma_2 \leftrightarrow \sigma_1, \sigma_4 \leftrightarrow \sigma_3)$, 
all logarithms of $\lambda$ are now seen to cancel. The resulting
integrand is sufficiently well-behaved around the boundaries
to be integrable [cf.\ \eq\nr{rho_Ij_final} and \se\ref{se:num}].

%
\section{Final result}
\la{se:final}

In order to collect together a final result, we remove 
redundant symmetries from the expression. It is suggested 
already by \eq\nr{calF} that the two substitutions
\be
  p\leftrightarrow \ko - q \; ; \quad p \leftrightarrow q
  \;,  \la{symmetries}
\ee
corresponding to reflections across the axes $q = \ko - p$
and $q=p$, respectively, may be helpful in this respect. 
Indeed, the first of these symmetries is manifest in the 
real corrections, and can
consequently be seen to transform the 
``coefficient functions'', 
\eqs\nr{coeff_first}--\nr{coeff_last}, into each other. 
The phase space distributions are in general not invariant
but now become symmetrized with respect to their indices
(concretely, a symmetry in
$(\sigma_1 \leftrightarrow \sigma_4, \sigma_2 \leftrightarrow \sigma_3)$
which was so far hidden becomes explicit).

A symmetry in $p\leftrightarrow q$ is not ``inherent'' to the 
expressions. It is useful to make this reflection, however, 
because it explicitly regulates principal value integrations
across $p-q = 0$. After these reflections, the integration 
range is as shown in \fig\ref{fig:limits}.

\begin{figure}[t]


\centerline{%
\epsfysize=7.5cm\epsfbox{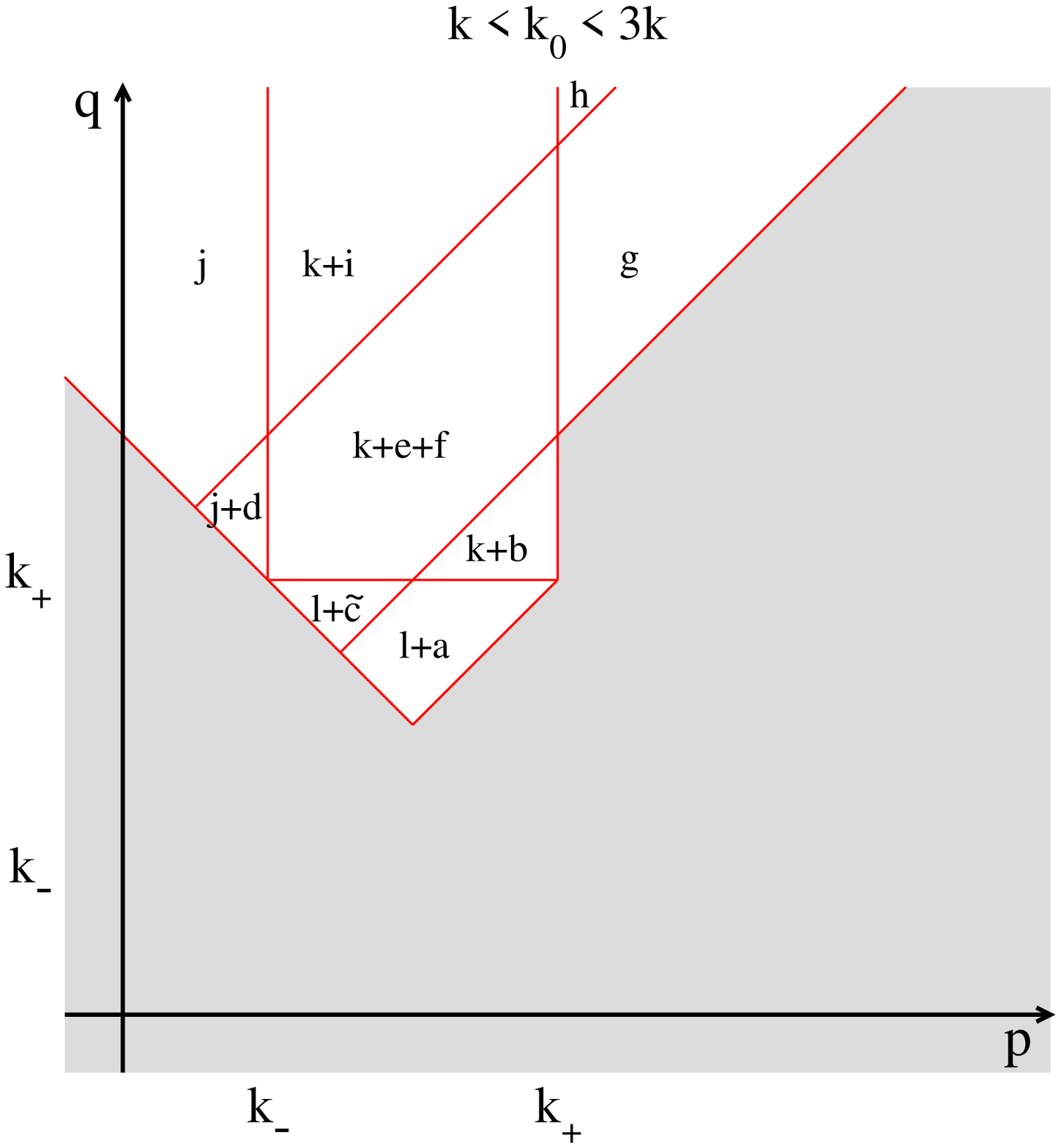}%
~~\epsfysize=7.5cm\epsfbox{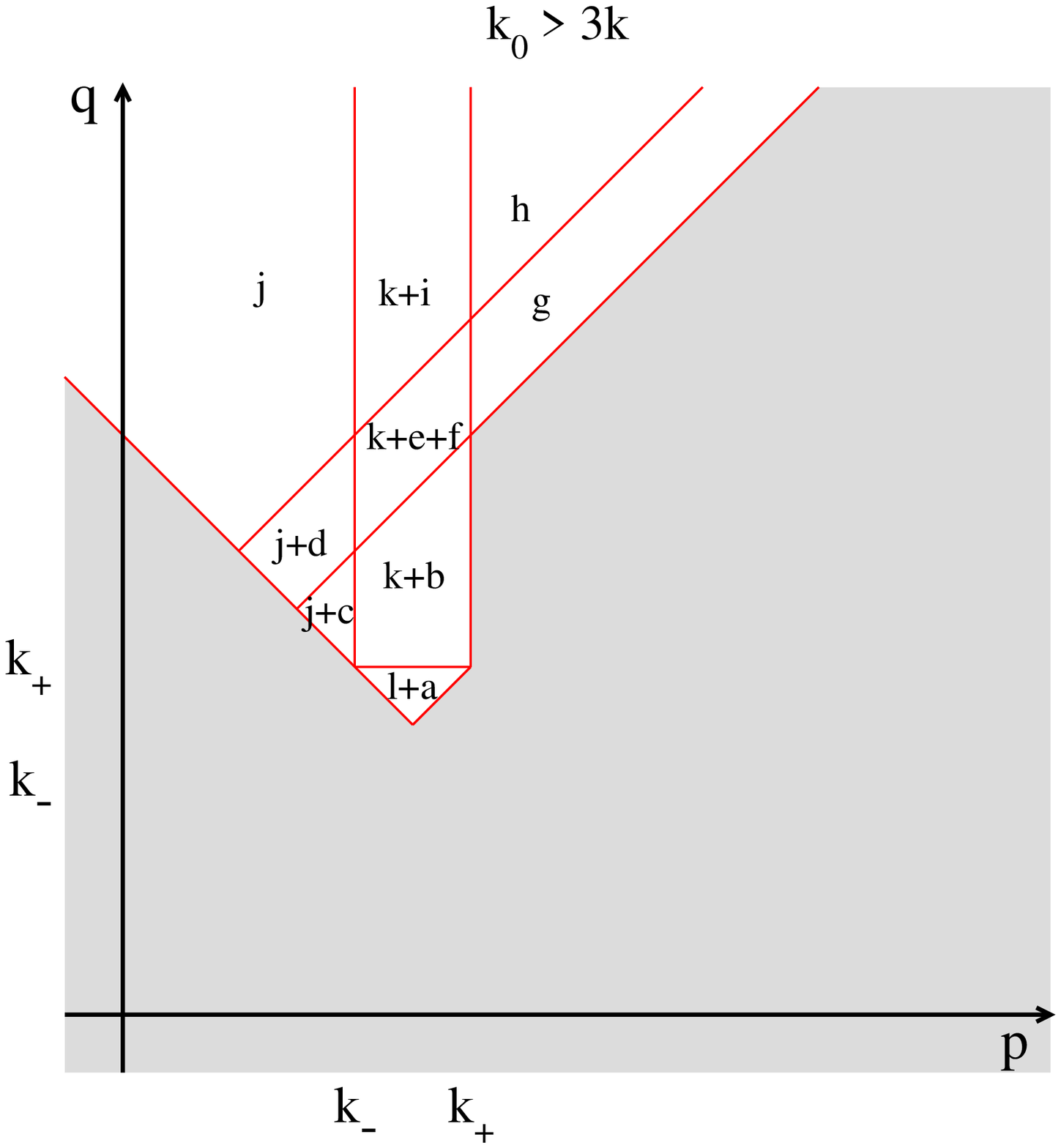}%
}

\caption[a]{\small 
 Integration ranges in the ($p,q$)-plane for the
 final answer, \eq\nr{rho_Ij_final}. The delimiting curves are 
 $
  p = \km 
 $,   
 $
  p = \kp 
 $,   
 $
 q = p
 $, 
 $
  q = p + \km 
 $, 
 $
  q = p + \kp 
 $, 
 $
  q = \ko - p
 $, as well as
 $
 q= \kp
 $. 
 }
\la{fig:limits}
\end{figure}

To present a final result, we undertake one more rewriting
of the phase space distributions. By making use of 
\eqs\nr{n_id_x}, \nr{n_identity}, the structures appearing in 
the real corrections are expressed as
\be
 n^{ }_{\sigma_4}(\ko -p)n^{ }_{\sigma_2}(q) 
 n^{ }_{\sigma_5}(p-q) = 
 n^{ }_{\sigma_4}(\ko -p) n^{ }_{\sigma_1}(p) 
 \Bigl\{ 
 \Bigl[
  \fr12 + n^{ }_{\sigma_2}(q) 
 \Bigr]
   - 
 \Bigl[
   \fr12 + n^{ }_{\sigma_5}(q-p)
 \Bigr]
 \Bigr\}
 \;,
\ee
where $n^{ }_{\sigma_5}$ has been taken with a positive argument. 
We then introduce the ``weight functions''
\ba
 \omega_1 & \equiv &  
   n^{ }_{\sigma_4}(\ko -p) n^{ }_{\sigma_1}(p) 
 \Bigr] \Bigl[ \fr12 + n^{ }_{\sigma_5}(q-p) \Bigr]
  + ({\scriptstyle \sigma_1,\sigma_4,\sigma_5})
  + ({\scriptstyle \sigma_2,\sigma_3,\sigma_5})
  + ({\scriptstyle \sigma_3,\sigma_2,\sigma_5})
 \;, \la{w1} \hspace*{5mm} \\
 \omega_2 & \equiv &  
   n^{ }_{\sigma_4}(\ko -q) n^{ }_{\sigma_1}(q) 
  \Bigl[ \fr12 + n^{ }_{\sigma_5}(q-p) \Bigr]
  + ({\scriptstyle \sigma_1,\sigma_4,\sigma_5})
  + ({\scriptstyle \sigma_2,\sigma_3,\sigma_5})
  + ({\scriptstyle \sigma_3,\sigma_2,\sigma_5})
 \;, \la{w2} \hspace*{5mm} \\
 \omega_3 & \equiv &  
   n^{ }_{\sigma_4}(\ko -p) n^{ }_{\sigma_1}(p) 
  \Bigl[ \fr12 + n^{ }_{\sigma_2}(q) \Bigr]
  + ({\scriptstyle \sigma_1,\sigma_4,\sigma_3})
  + ({\scriptstyle \sigma_2,\sigma_3,\sigma_4})
  + ({\scriptstyle \sigma_3,\sigma_2,\sigma_1})
 \;, \la{w3} \hspace*{5mm} \\
 \omega_4 & \equiv &  
   n^{ }_{\sigma_4}(\ko -q) n^{ }_{\sigma_1}(q) 
  \Bigl[ \fr12 + n^{ }_{\sigma_2}(p) \Bigr]
  + ({\scriptstyle \sigma_1,\sigma_4,\sigma_3})
  + ({\scriptstyle \sigma_2,\sigma_3,\sigma_4})
  + ({\scriptstyle \sigma_3,\sigma_2,\sigma_1})
 \;. \la{w4} \hspace*{5mm} 
\ea
To characterize the ``coefficient functions'' 
it is convenient to define the ratios
\ba
 && 
 \chi^{ }_- \; \equiv \; \frac{q-\km}{q-\ko} \;, \quad
 \chi^{ }_+ \; \equiv \; \frac{q-\kp}{q-\ko} \;, \quad
 \chi^{ }_0 \; \equiv \; \frac{q}{q-\ko} \;, \\
 && 
 \pi^{ }_- \; \equiv \; \frac{p-\km}{p-\ko} \;, \quad
 \pi^{ }_+ \; \equiv \; \frac{p-\kp}{p-\ko} \;, \quad
 \pi^{ }_0 \; \equiv \; \frac{p}{p-\ko}
 \;.
\ea 
Then the full result 
becomes 
\ba
 && \hspace*{-1cm} 
 \frac{(4\pi)^4 k\, \rho^{ }_{\mathcal{I}^{ }_\rmii{j}}(\mathcal{K}) }
 {\pi \mathcal{K}^2 n^{-1}_{\sigma_0}(\ko)}  =  
 \nn &&
 \int^{ }_{\Omega^{ }_\rmii{l+a}} \!   
 \frac{{\rm d}p\,{\rm d}q}{q-p}\biggl\{
   (\omega_1 - \omega_2)
   \ln\biggl|
      \frac{\pi^{ }_-\pi^{ }_+}{\chi^{ }_- \chi^{ }_+}
   \biggr|
   + \, \omega_3
   \ln\biggl|
      \frac{\chi^2_0}{\chi^{ }_- \chi^{ }_+ \pi^{ }_-\pi^{ }_+ }
   \biggr|
   + \omega_4
   \ln\biggl|
      \frac{\chi^{ }_- \chi^{ }_+ \pi^{ }_-\pi^{ }_+ }{\pi^2_0}
   \biggr|
 \biggr\}
 \nn & + & 
 \int^{ }_{\Omega^{ }_\rmii{k+b}} \!   
 \frac{{\rm d}p\,{\rm d}q}{q-p}\biggl\{
   \omega_1
   \ln\biggl|
      \frac{\chi^{ }_0 \pi^{ }_+}{\chi^{ }_-\chi^2_+}
   \biggr|
   + \, (\omega_2 + \omega_4) 
   \ln\biggl|
      \frac{\chi^{ }_-\pi^{ }_+}{\pi^{ }_0}
   \biggr|
   + \omega_3
   \ln\biggl|
      \frac{\chi^{ }_0}{\chi^{ }_-\pi^{ }_+}
   \biggr|
 \biggr\}
 \nn & + & 
 \int^{ }_{\Omega^{ }_\rmii{j+c}} \!   
 \frac{{\rm d}p\,{\rm d}q}{q-p}\biggl\{
   (\omega_1+\omega_2-\omega_3+\omega_4)
   \ln\biggl|
      \frac{\chi^{ }_-\pi^{ }_+}{\chi^{ }_+\pi^{ }_-}
   \biggr|
 \biggr\}
 \nn & + & 
 \int^{ }_{\Omega^{ }_\rmii{l+$\tilde{c}$} } \!   
 \frac{{\rm d}p\,{\rm d}q}{q-p}\biggl\{
   \omega_1
   \ln\biggl|
      \frac{\pi^{ }_-\pi^{ }_+}{\chi^{ }_-\chi^{ }_+}
   \biggr|
   + \, \omega_2 
   \ln\biggl|
      \frac{\pi^{ }_0 \chi^{ }_+}{\pi^{ }_-\pi^2_+}
   \biggr|
   + \omega_3
   \ln\biggl|
      \frac{\chi^2_0}{\chi^{ }_- \chi^{ }_+ \pi^{ }_-\pi^{ }_+ }
   \biggr|
   + \omega_4
   \ln\biggl|
      \frac{\chi^{ }_+ \pi^{ }_-}{ \pi^{ }_0 }
   \biggr|
 \biggr\}
 \nn & + & 
 \int^{ }_{\Omega^{ }_\rmii{j+d}} \!   
 \frac{{\rm d}p\,{\rm d}q}{q-p}\biggl\{
   (\omega_1-\omega_3)
   \ln\biggl|
      \frac{\chi^{ }_- \pi^{ }_+}{ \chi^{ }_+\pi^{ }_- }
   \biggr|
   + \, (\omega_2 + \omega_4) 
   \ln\biggl|
      \frac{\pi^{ }_0}{\chi^{ }_+ \pi^{ }_- }
   \biggr|
 \biggr\}
 \nn & + & 
 \int^{ }_{\Omega^{ }_\rmii{k+e+f}} \!   
 \frac{{\rm d}p\,{\rm d}q}{q-p}\biggl\{
   \omega_1
   \ln\biggl|
      \frac{\chi^{ }_0 \pi^{ }_+ }{\chi^{ }_- \chi^2_+  }
   \biggr|
   +\, \omega_3
   \ln\biggl|
      \frac{\chi^{ }_0}{\chi^{ }_- \pi^{ }_+  }
   \biggr|
 \biggr\}
 \nn & + & 
 \int^{ }_{\Omega^{ }_\rmii{g}} \!   
 \frac{{\rm d}p\,{\rm d}q}{q-p}\biggl\{
  (\omega_2 + \omega_4) 
   \ln\biggl|
      \frac{\pi^{ }_0}{\chi^{ }_- \pi^{ }_+  }
   \biggr|
 \biggr\}
 \nn & + & 
 \int^{ }_{\Omega^{ }_\rmii{h}} \!   
 \frac{{\rm d}p\,{\rm d}q}{q-p}\biggl\{
  (\omega_2 + \omega_4) 
   \ln\biggl|
      \frac{\chi^{ }_+ \pi^{ }_-}{\chi^{ }_- \pi^{ }_+  }
   \biggr|
 \biggr\}
 \nn & + & 
 \int^{ }_{\Omega^{ }_\rmii{k+i}} \!   
 \frac{{\rm d}p\,{\rm d}q}{q-p}\biggl\{
   \omega_1
   \ln\biggl|
      \frac{\chi^{ }_0 \pi^{ }_+}{\chi^{ }_- \chi^2_+  }
   \biggr|
   + \, (\omega_2 + \omega_4) 
   \ln\biggl|
      \frac{\chi^{ }_+ \pi^{ }_-  } {\pi^{ }_0}
   \biggr|
   + \omega_3
   \ln\biggl|
      \frac{\chi^{ }_0 }{\chi^{ }_- \pi^{ }_+  }
   \biggr|
 \biggr\}
 \nn & + & 
 \int^{ }_{\Omega^{ }_\rmii{j}} \!   
 \frac{{\rm d}p\,{\rm d}q}{q-p}\biggl\{
   (\omega_1-\omega_3)
   \ln\biggl|
      \frac{\chi^{ }_- \pi^{ }_+}{\chi^{ }_+ \pi^{ }_-  }
   \biggr|
 \biggr\}
 \;, \la{rho_Ij_final}
\ea
where the $\Omega$'s denote different domains as
labelled in \fig\ref{fig:limits}. Note that only one among
the ranges $\Omega_\rmi{j+c}$ and 
$\Omega_\rmi{l+$\tilde{\rm c}$}$ gets realized at a time.

%
\section{Numerical evaluation}
\la{se:num}

\begin{figure}[t]


\centerline{%
 \epsfysize=7.5cm\epsfbox{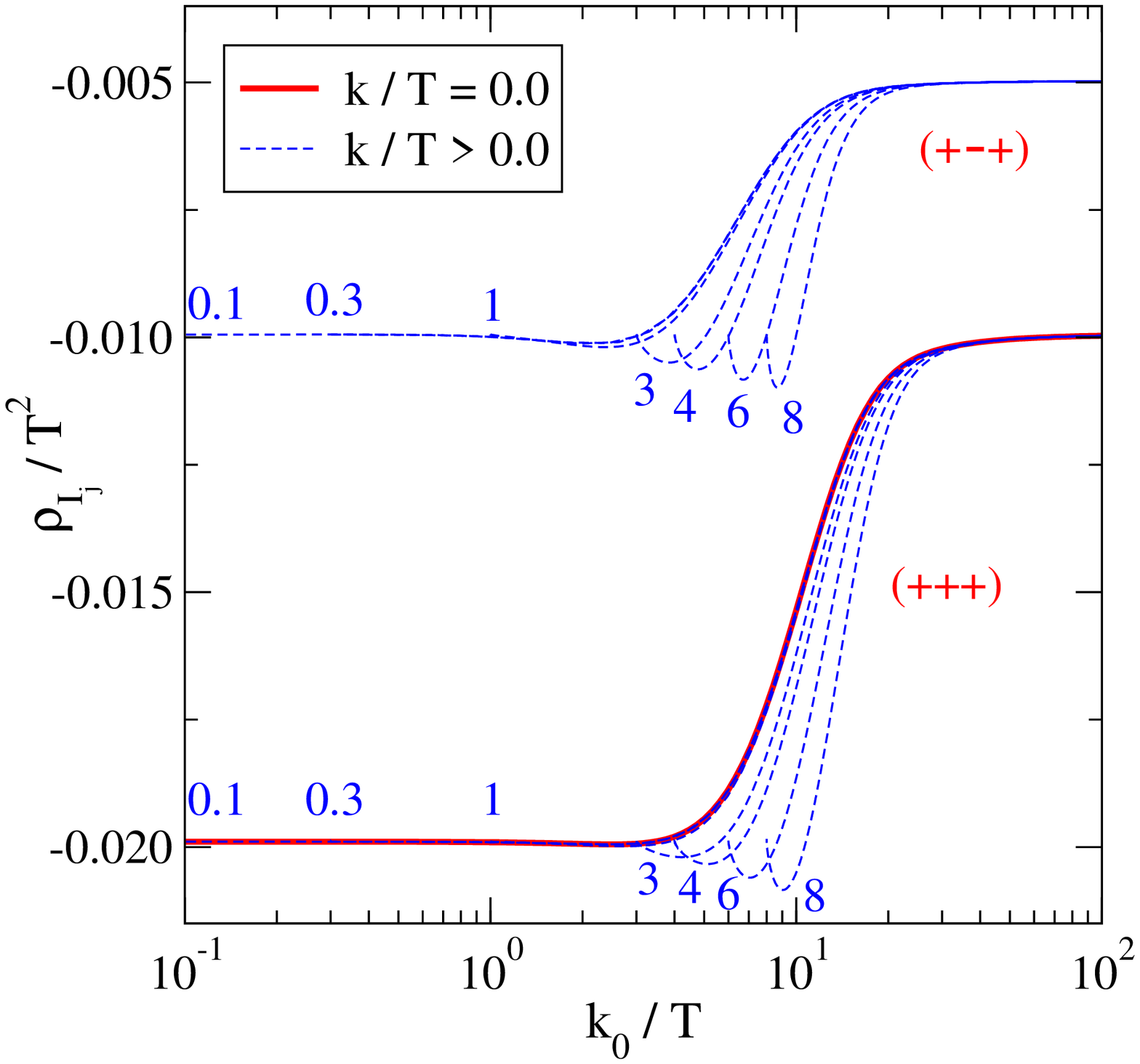}%
~~~\epsfysize=7.5cm\epsfbox{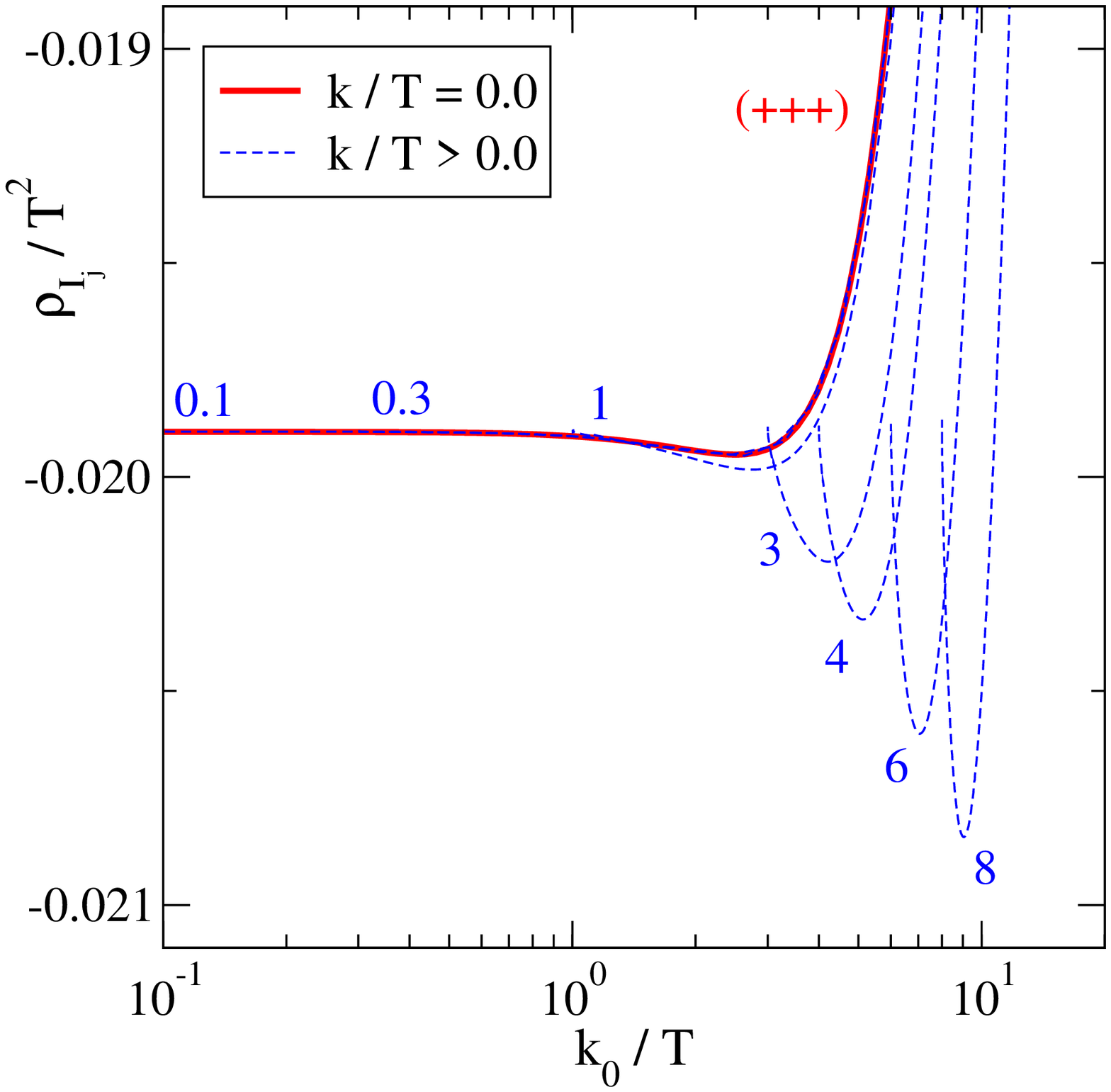}
}

\caption[a]{\small
Left: A comparison of the spectral function at non-zero momentum
($k/T =$ 0.1, 0.3, 1, 3, 4, 6, 8 as indicated in the figure)
with the zero-momentum limit from \eq(A.57) of ref.~\cite{bulk_wdep}, for 
$(\sigma_1 \sigma_4 \sigma_5)$ = ($+$\,$+$\,$+$). 
For the case $(\sigma_1 \sigma_4 \sigma_5)$ = ($+$\,$-$\,$+$)
only non-zero momenta are shown. 
In both cases we have restricted to $\ko  \ge  k + 0.001 T$. 
Right: A magnification of the case 
$(\sigma_1 \sigma_4 \sigma_5)$ = ($+$\,$+$\,$+$). 
}

\la{fig:numerics}
\end{figure}

The expression in \eq\nr{rho_Ij_final} is finite and can be 
evaluated numerically. In \fig\ref{fig:numerics} the outcome
is compared with its limiting value at $k = 0$, determined for 
the case $(\sigma_1 \sigma_4 \sigma_5)$ = 
($+$\,$+$\,$+$) in ref.~\cite{bulk_wdep}.
The results are seen to agree for $\ko \gg k$;
in fact,  
even when this inequality is not satisfied, the zero-momentum 
limit yields a surprisingly good approximation of the full result.  

It is important to realize that 
the spectral
function $\rho^{ }_{\mathcal{I}^{ }_\rmii{j}}$ is in general  
non-trivial in the vicinity of the light cone, 
cf.\ \fig\ref{fig:numerics}, rather than vanishing 
as $\sim \mathcal{K}^2$ as one could expect from dimensional
reasons at zero temperature. 
This fact leads 
ultimately to the breakdown of the loop expansion for 
$\mathcal{K}^2 \lsim (g T)^2$ 
[2-loop diagrams may give a larger contribution than 1-loop ones
despite the overall suppression by $g^2$], 
and leads to the need to resum a set of diagrams for 
obtaining the correct result in the ultrarelativistic regime.  

\begin{figure}[t]


\centerline{%
 \epsfysize=7.5cm\epsfbox{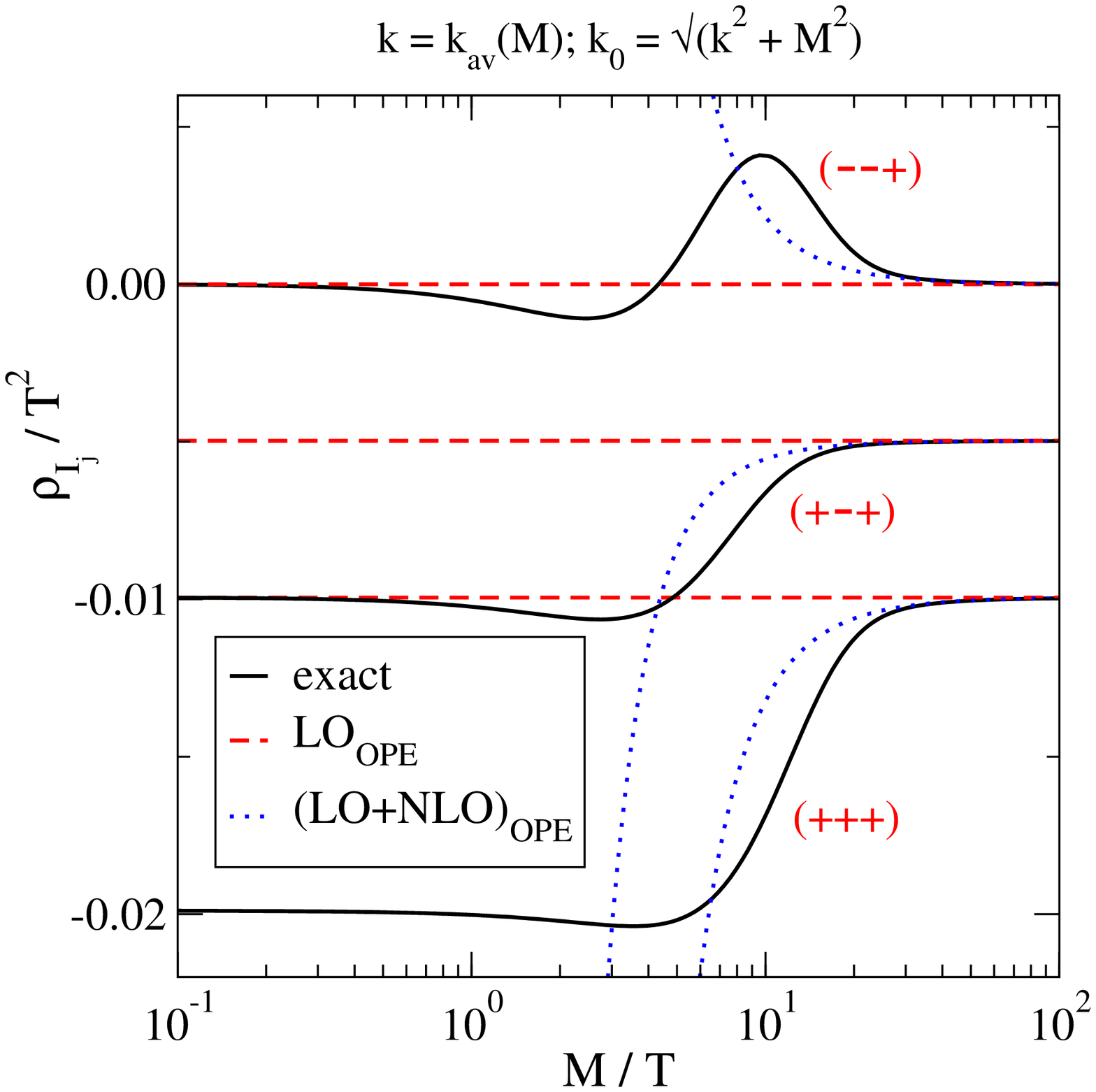}%
~~~\epsfysize=7.5cm\epsfbox{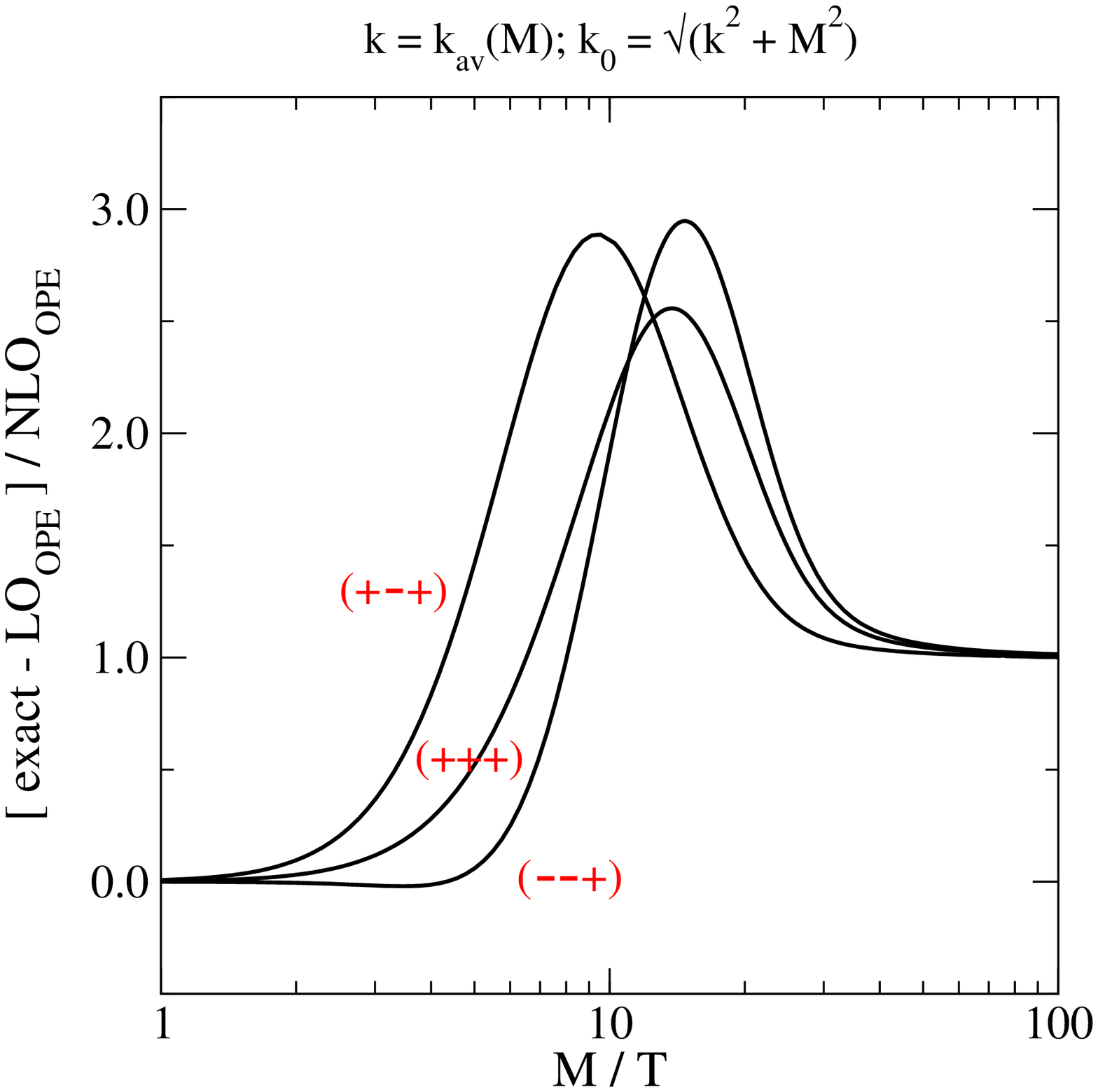}
}

\caption[a]{\small
Left: Spectral functions for various statistics
$(\sigma_1 \sigma_4 \sigma_5)$, as a function
of $M^2 \equiv \mathcal{K}^2$, with $k \equiv k_\rmi{av}^{ }$
determined from \eq\nr{kav}. The results are compared with 
two orders of OPE asymptotics from \eq\nr{asymptotics}.  
Right: The ratio 
[exact - LO$^{ }_\rmii{OPE}$]/NLO$^{ }_\rmii{OPE}$, showing
that numerical results agree with the asymptotic
ones from \eq\nr{asymptotics} for $M \gsim 30 T$. 
}

\la{fig:ope}
\end{figure}

Another comparison can be made with the non-relativistic asymptotics, 
determined in ref.~\cite{nonrel}. The leading term, proportional to 
$\mathcal{K}^2$, corresponds to the zero-temperature limit and vanishes
for the spectral function in question. The two first non-zero terms 
read 
\be
  \rho^{ }_{\mathcal{I}^{ }_\rmii{j}} =  
  -\frac{1}{16\pi} \int_{p} 
  \biggl\{ 
  \frac{\sum_{i=1}^{4} n^{ }_{\sigma_i}  +
  2 n^{ }_{\sigma_5}}{p}
 + \, 
 p \,
 \biggl[
  \frac{22 }{3} \sum_{i=1}^{4} n^{ }_{\sigma_i}
  + 4 n^{ }_{\sigma_5} 
 \biggr] \frac{\ko^2 + k^2/3}{\mathcal{K}^4} \biggr\}
 + \rmO\Bigl(\frac{T^6}{\mathcal{K}^4}\Bigr)
 \;, \la{asymptotics}
\ee
where 
\be
 \int_p \frac{\nB{}}{p} = \frac{T^2}{12}
 \;, \quad
 \int_p \frac{\nF{}}{p} = \frac{T^2}{24}
 \;, \quad
 \int_p p\, \nB{} = \frac{\pi^2 T^4}{30}
 \;, \quad
 \int_p p\, \nF{} = \frac{7 \pi^2 T^4}{240}
 \;.
\ee
In \fig\ref{fig:ope} the two 
orders shown in \eq\nr{asymptotics}
are referred to as ``LO$^{ }_\rmii{OPE}$'' 
and ``NLO$^{ }_\rmii{OPE}$'', respectively. In order to carry out 
the comparison in a somewhat realistic setting,  
we  introduce a phenomenological average momentum through  
\be
 k^2_\rmi{av}(M) \equiv \frac{\int_0^\infty \! {\rm d}k \, k^4 
 \exp(-\frac{\sqrt{k^2 + M^2}}{T}) }{\int_0^\infty \! {\rm d}k \, k^2 
 \exp(-\frac{\sqrt{k^2 + M^2}}{T})} = 
 \frac{3 M T K^{ }_3(\fr{M}{T})}{K^{ }_2(\fr{M}{T})}
 \;. \la{kav}
\ee
This should be understood just as a rough guideline; for instance 
we have employed a Boltzmann weight so that the same value can be 
used for any statistics. In any case, as can clearly be seen
in \fig\ref{fig:ope}(right), the correct limits are 
reached for all statistics considered, if only quite
deep in the non-relativistic regime.

%
\section{Summary and outlook}
\la{se:concl}

The purpose of this paper has been to suggest
a general strategy for determining 2-loop thermal spectral functions
at non-zero energy and momentum in the rest frame of 
a heat bath. As has been demonstrated
with the example of the most complicated ``master'' structure, 
the result can be reduced to a convergent 2-dimensional 
integral, \eq\nr{rho_Ij_final}, 
within a domain shown in \fig\ref{fig:limits}. 
For other master spectral functions, the domain remains
the same but the ``coefficient functions''  change; in addition,  
due to less symmetry, the  ``weight functions'' 
split up into a larger set of independent ones
(the weight functions are defined as polynomials
of the phase space distributions incorporating 
all the temperature dependence).\footnote{%
  It should be mentioned that for some of the simpler masters it is 
  not necessary to make use of the full formalism introduced 
  in the present paper, however if other tricks fail one can
  always resort to it. 
 }

The specific master spectral function studied, defined
by \eqs\nr{def_Ij}, \nr{cut}, is peculiar in that it vanishes
in the zero-temperature limit (this comes about through a 
complete cancellation of real and virtual corrections, and 
is reproduced by our numerical results). 
Therefore it is natural to express it as 
$\rho^{ }_{\mathcal{I}^{ }_\rmii{j}}
 =  T^2 \phi({\kp} / {T}, {\km} / {T})$, 
where $k^{}_\pm \equiv (\ko\pm k)/2$ and $\phi$ is 
a dimensionless function. 
The function $\phi$ has a finite limiting value in the non-relativistic 
limit $\kp,\km \gg \pi T$, but a non-trivial structure in the relativistic
regime $\kp,\km \sim \pi T$, cf.\ \fig\ref{fig:numerics}. 
(In the regime $\km \ll \pi T$ the naive loop expansion 
of thermal field theory breaks down and needs to be resummed
through effective field theory techniques.) 

Several extensions of the current 
investigation can be envisaged. 
The most obvious challenges are to work out similar results for
the other master structures appearing in \eqs\nr{dilepton}, \nr{neutrino}
and then to compile results for the physical observables discussed
in \se\ref{se:reduce}. It might also be interesting to extend 
the results to a situation where some of the propagators are 
massive; this would be relevant for the cosmological 
applications reviewed in ref.~\cite{numsm}. (As has been 
demonstrated with a particular non-zero mass here, it may be possible to 
reduce the result to a 2-dimensional integral even in 
the presence of masses.) 
Perhaps it would
be nice to understand analytically the behaviour in the regime
$\mathcal{K}^2 \ll (\pi T)^2$.
In addition the question could be posed whether, possibly with the 
price of introducing one further integration variable, the 
final result in \eq\nr{rho_Ij_final} could be cast in a more
compact and transparent form. Last but not least, the computation presented 
involved a fair amount of error-prone hand work, so that an 
independent crosscheck, perhaps involving
other integration variables and/or another intermediate 
infrared regulator, would be more than welcome. 

%
\section*{Acknowledgements}

I am grateful to D.~B\"odeker and Y.~Schr\"oder for helpful discussions
and suggestions. 
This work was partly supported by the Swiss National Science Foundation
(SNF) under grant 200021-140234.

%
\appendix
\renewcommand{\thesection}{Appendix~\Alph{section}}
\renewcommand{\thesubsection}{\Alph{section}.\arabic{subsection}}
\renewcommand{\theequation}{\Alph{section}.\arabic{equation}}



\begin{thebibliography}{99}

\bibitem{simon}
  S.~Caron-Huot,
  {\it Asymptotics of thermal spectral functions,}
  Phys.\ Rev.\  D {79} (2009) 125009
  [0903.3958].

\bibitem{amypre}
  P.B.~Arnold, G.D.~Moore and L.G.~Yaffe,
  {\em Photon emission from ultrarelativistic plasmas,}
  JHEP {11} (2001) 057
  [hep-ph/0109064].

\bibitem{amy}
  P.B.~Arnold, G.D.~Moore and L.G.~Yaffe,
  {\em Photon emission from quark gluon plasma: 
  Complete leading order results,}
  JHEP {12} (2001) 009
  [hep-ph/0111107].

\bibitem{amypost}
  P.B.~Arnold, G.D.~Moore and L.G.~Yaffe,
  {\em Photon and gluon emission in relativistic plasmas,}
  JHEP {06} (2002) 030
  [hep-ph/0204343].

\bibitem{bp}
  E.~Braaten, R.D.~Pisarski and T.-C.~Yuan,
  {\em Production of soft dileptons in the quark--gluon plasma,}
  Phys.\ Rev.\ Lett.\  {64} (1990) 2242.

\bibitem{ag}
  P.~Aurenche, F.~Gelis, G.D.~Moore and H.~Zaraket,
  {\em Landau-Pomeranchuk-Migdal resummation for dilepton production,}
  JHEP {12} (2002) 006
  [hep-ph/0211036].

\bibitem{mr}
  G.D.~Moore and J.-M.~Robert,
  {\em Dileptons, spectral weights, and conductivity 
  in the quark-gluon plasma,}
  hep-ph/0607172.

\bibitem{bb0}
  D.~Besak and D.~B\"odeker,
  {\em Hard Thermal Loops for Soft or Collinear External Momenta,}
  JHEP {05} (2010) 007
  [1002.0022].

\bibitem{bb1}
  A.~Anisimov, D.~Besak and D.~B\"odeker,
  {\it Thermal production of relativistic Majorana neutrinos: 
  Strong enhancement by multiple soft scattering,}
  JCAP {03} (2011) 042
  [1012.3784].

\bibitem{bb2}
  D.~Besak and D.~B\"odeker,
  {\it Thermal production of ultrarelativistic right-handed 
  neutrinos: Complete leading-order results,}
  JCAP {03} (2012) 029
  [1202.1288].

\bibitem{photonNLO}
  J.~Ghiglieri, J.~Hong, A.~Kurkela, E.~Lu, G.D.~Moore and D.~Teaney,
  {\em Next-to-leading order thermal photon production 
  in a weakly coupled quark-gluon plasma,}
  1302.5970.

\bibitem{aarts}
  G.~Aarts and J.M.~Mart{\'i}nez Resco,
  {\em Continuum and lattice meson spectral functions 
  at nonzero momentum and high temperature,}
  Nucl.\ Phys.\ B {726} (2005) 93
  [hep-lat/0507004].

\bibitem{yanagida}
  M.~Fukugita and T.~Yanagida,
  {\em Baryogenesis without Grand Unification,}
  Phys.\ Lett.\  B {174} (1986) 45.

\bibitem{numsm}
  L.~Canetti, M.~Drewes, T.~Frossard and M.~Shaposhnikov,
  {\em Dark Matter, Baryogenesis and Neutrino Oscillations from  
  Right Handed Neutrinos,}
  1208.4607.

\bibitem{salvio}
  A.~Salvio, P.~Lodone and A.~Strumia,
  {\it Towards leptogenesis at NLO: 
  the right-handed neutrino interaction rate,}
  JHEP {08} (2011) 116
  [1106.2814].

\bibitem{nonrel}
  M.~Laine and Y.~Schr\"oder,
  {\em Thermal right-handed neutrino production rate
  in the non-relativistic regime,}
  JHEP {02} (2012) 068
  [1112.1205].

\bibitem{spectral1}
  R.~Baier, B.~Pire and D.~Schiff,
  {\it Dilepton production at finite temperature: 
  Perturbative treatment at order $\alpha_s$,}
  Phys.\ Rev.\  D {38} (1988) 2814.

\bibitem{spectral2}
  Y.~Gabellini, T.~Grandou and D.~Poizat,
  {\it Electron-positron annihilation in thermal QCD,}
  Annals Phys.\  {202} (1990) 436.

\bibitem{spectral3}
  T.~Altherr and P.~Aurenche,
  {\it Finite temperature QCD corrections to lepton-pair 
  formation in a quark-gluon plasma,}
  Z.\ Phys.\  C {45} (1989) 99.

\bibitem{bulk_wdep}
  M.~Laine, A.~Vuorinen and Y.~Zhu,
  {\em Next-to-leading order thermal spectral functions
  in the perturbative domain,}
  JHEP {09} (2011)  084
  [1108.1259].

\bibitem{shear_wdep}
  Y.~Zhu and A.~Vuorinen,
  {\em The shear channel spectral function in hot Yang-Mills theory,}
  JHEP {03} (2013) 002
  [1212.3818].

\bibitem{cuniberti}
  G.~Cuniberti, E.~De Micheli and G.A.~Viano,
  {\em Reconstructing the thermal Green functions at real times from those at
  imaginary times,}
  Commun.\ Math.\ Phys.\  {216} (2001) 59
  [cond-mat/0109175].

\bibitem{analytic}
  Y.~Burnier, M.~Laine and L.~Mether,
  {\it A test on analytic continuation of thermal imaginary-time data,}
  Eur.\ Phys.\ J.\  C {71} (2011) 1619
  [1101.5534].

\bibitem{cond}
  Y.~Burnier and M.~Laine,
 {\em Towards flavour diffusion coefficient and 
  electrical conductivity without ultraviolet contamination,}
  Eur.\ Phys.\ J.\ C {72} (2012) 1902
  [1201.1994].

\bibitem{akr}
  A.K.~Rajantie,
  {\em Feynman diagrams to three loops in three-dimensional field theory,}
  Nucl.\ Phys.\ B {480} (1996) 729
  [Erratum-ibid.\ B {513} (1998) 761]
  [hep-ph/9606216].

\bibitem{db}
  D.J.~Broadhurst, J.~Fleischer and O.V.~Tarasov,
  {\em Two-loop two-point functions with masses: 
  Asymptotic expansions and Taylor series, in any dimension,}
  Z.\ Phys.\ C {60} (1993) 287
  [hep-ph/9304303].

\bibitem{vs}
  V.A.~Smirnov, 
  {\em Analytic Tools for Feynman Integrals}
  (Springer Verlag, Berlin, 2012). 

\bibitem{bulk_rdep}
  M.~Laine, M.~Veps\"al\"ainen and A.~Vuorinen,
  {\em Intermediate distance correlators in hot Yang-Mills theory,}
  JHEP {12} (2010) 078
  [1011.4439].

\bibitem{gh}
  B.~Garbrecht, F.~Glowna and M.~Herranen,
  {\em Right-Handed Neutrino Production at Finite Temperature: 
  Radiative Corrections, Soft and Collinear Divergences,}
  1302.0743.

\bibitem{dilepton1}
  L.D.~McLerran and T.~Toimela,
  {\it Photon and Dilepton Emission from the 
   Quark-Gluon Plasma: Some General Considerations,}
  Phys.\ Rev.\ D {31} (1985) 545.

\bibitem{dilepton2}
  H.A.~Weldon,
  {\it Reformulation of Finite Temperature Dilepton Production,}
  Phys.\ Rev.\ D {42} (1990) 2384.

\bibitem{dilepton3}
  C.~Gale and J.I.~Kapusta,
  {\em Vector dominance model at finite temperature,}
  Nucl.\ Phys.\ B {357} (1991) 65.

\bibitem{GVtau}
  Y.~Burnier and M.~Laine,
  {\em Massive vector current correlator in thermal QCD,}
  JHEP {11} (2012) 086
  [1210.1064].

\bibitem{bulk_ope}
  M.~Laine, M.~Veps\"al\"ainen and A.~Vuorinen,
  {\it Ultraviolet asymptotics of scalar and pseudoscalar
  correlators in hot Yang-Mills theory,}
  JHEP {10} (2010) 010
  [1008.3263].

\bibitem{shear_ope}
  Y.~Schr\"oder, M.~Veps\"al\"ainen, A.~Vuorinen and Y.~Zhu,
  {\em The ultraviolet limit and sum rule for the shear correlator 
  in hot Yang-Mills theory,}
  JHEP {12} (2011) 035
  [1109.6548].

\bibitem{kit}
  M.F.~Zoller and K.G.~Chetyrkin,
  {\em OPE of the energy-momentum tensor correlator in massless QCD,}
  JHEP {12} (2012) 119
  [1209.1516].

\bibitem{selfE}
  M.~Laine,
  {\em Thermal right-handed neutrino self-energy 
  in the non-relativistic regime,}
  1209.2869.

\bibitem{hw}
  H.A.~Weldon,
  {\it Effective fermion masses of $\rmO(gT)$ 
  in high-temperature gauge theories
  with exact chiral invariance,}
  Phys.\ Rev.\ D {26} (1982) 2789.

\bibitem{weldon}
  H.A.~Weldon,
  {\em Simple Rules for Discontinuities in Finite Temperature Field Theory,}
  Phys.\ Rev.\ D {28} (1983) 2007.

\bibitem{nlo}
  Y.~Burnier, M.~Laine and M.~Veps\"al\"ainen,
  {\em Heavy quark medium polarization at next-to-leading order,}
  JHEP {02} (2009) 008
  [0812.2105].


\bibitem{bk}
  E.~Byckling and K.~Kajantie, 
  {\em Particle Kinematics}
  (John~Wiley, New York, 1973). 


\end{thebibliography}
\end{document}